\pdfoutput=1 
\documentclass[cernpreprint,english,USenglish]{na61doc}

\usepackage{lineno, blindtext}
\usepackage{amsmath}% loads amstext, amsbsy, amsopn but not amssymb
\usepackage{enumerate}
\usepackage{placeins}
\usepackage{appendix}
\usepackage[LY1]{fontenc}
\usepackage[utf8]{inputenc}
\usepackage{chngcntr}
\usepackage{microtype}
\usepackage{lineno}
\usepackage{color}
\usepackage{epstopdf}
\usepackage{colortbl}
\definecolor{darkred}{rgb}{0.5,0,0}
\definecolor{darkblue}{rgb}{0,0,0.5}
\definecolor{firebrick}{rgb}{0.75,0.125,0.125}
\definecolor{darkgreen}{rgb}{0,0.5,0}

\usepackage{amsmath}
\usepackage{graphicx}
\usepackage{booktabs}
\usepackage{tabularx}
\usepackage{dcolumn}% Align table columns on decimal point
\usepackage{bm}% bold math
\usepackage{xspace} % to add space after command
\usepackage{longtable}% needed for data tables
\usepackage{caption} % needed to fix caption width?
\usepackage{url}

\usepackage[colorinlistoftodos]{todonotes}

\newcommand{\GeVc}{\mbox{GeV/$c$}\xspace}
\newcommand{\GeVcc}{\mbox{GeV/$c^2$}\xspace}
\newcommand{\MeVcc}{\mbox{MeV/$c^2$}\xspace}
\newcommand{\piC}{\ensuremath{\pi^+ + \textup{C at 60 } \GeVc} \xspace}
\newcommand{\piBe}{\ensuremath{\pi^+ + \textup{Be at 60 } \GeVc}\xspace}
\newcommand{\pip}{\ensuremath{\pi^+}\xspace}
\newcommand{\pim}{\ensuremath{\pi^-}\xspace}
\newcommand{\pipm}{\ensuremath{\pi^\pm}\xspace}
\newcommand{\kp}{\ensuremath{K^+}\xspace}
\newcommand{\km}{\ensuremath{K^-}\xspace}

\newcommand{\mupm}{\ensuremath{\mu^\pm}\xspace}
\newcommand{\epl}{\ensuremath{e^+}\xspace}

\newcommand{\kos}{\ensuremath{K^0_{\textup{S}}}\xspace}

\newcommand{\lam}{\ensuremath{\Lambda}\xspace}
\newcommand{\alam}{\ensuremath{\overline{\Lambda}}\xspace}

\ShineTitle{Measurements of hadron production in \boldmath{$\pi^{+}$} + C and \\ \boldmath{$\pi^{+}$} + Be interactions at 60\,\GeVc}
\PreprintIdNumber{CERN-EP-2019-198}

\begin{document}
\maketitle

%***********************************************************************************

%\linenumbers
\newpage 
{\Large The \NASixtyOne Collaboration}
\bigskip

\noindent
A.~Aduszkiewicz$^{\,15}$,
E.V.~Andronov$^{\,21}$,
T.~Anti\'ci\'c$^{\,3}$,
V.~Babkin$^{\,19}$,
M.~Baszczyk$^{\,13}$,
S.~Bhosale$^{\,10}$,
A.~Blondel$^{\,23}$,
M.~Bogomilov$^{\,2}$,
A.~Brandin$^{\,20}$,
A.~Bravar$^{\,23}$
% ...

\noindent
% ...
$^{2}$~Faculty of Physics, University of Sofia, Sofia, Bulgaria\\
$^{3}$~Ru{\dj}er Bo\v{s}kovi\'c Institute, Zagreb, Croatia\\
% ...
$^{10}$~Institute of Nuclear Physics, Polish Academy of Sciences, Cracow, Poland\\
% ...
$^{13}$~AGH - University of Science and Technology, Cracow, Poland\\
% ...
$^{15}$~University of Warsaw, Warsaw, Poland\\
% ...
$^{19}$~Joint Institute for Nuclear Research, Dubna, Russia\\
$^{20}$~National Research Nuclear University (Moscow Engineering Physics Institute), Moscow, Russia\\
$^{21}$~St. Petersburg State University, St. Petersburg, Russia\\
% ...
$^{23}$~University of Geneva, Geneva, Switzerland\\
% ...

\date{\today}

\begin{abstract}
	Precise knowledge of hadron production rates in the generation of neutrino beams is necessary for accelerator-based neutrino experiments to achieve their physics goals. NA61/SHINE, a large-acceptance hadron spectrometer, has recorded hadron+nucleus interactions relevant to ongoing and future long-baseline neutrino experiments at Fermi National Accelerator Laboratory. This paper presents three analyses of interactions of 60 \GeVc $\pi^+$ with thin, fixed carbon and beryllium targets. Integrated production and inelastic cross sections were measured for both of these reactions. In an analysis of strange, neutral hadron production, differential production multiplicities of \kos, \lam and \alam were measured. Lastly, in an analysis of charged hadron production, differential production multiplicities of \pip, \pim, \kp, \km and protons were measured. These measurements will enable long-baseline neutrino experiments to better constrain predictions of their neutrino flux in order to achieve better precision on their neutrino cross section and oscillation measurements.
\end{abstract}

\section{Introduction}

The NA61 or SPS Heavy Ion and Neutrino Experiment (SHINE)~\cite{na61detector} has a broad physics program that includes heavy ion physics, cosmic ray physics and neutrino physics. Accelerator-generated neutrino beams rely on beams of high energy protons which are directed towards a fixed target. The interactions of these protons result in secondary hadrons (especially pion, kaons, protons, neutrons and lambdas), some of which decay to produce the beam of neutrinos.  As most neutrino beam lines use targets that are an interaction length or longer in length, many of the secondary hadrons can re-interact inside the target and other beam material (such as the decay pipe walls or material of the focusing horns). Thus, it is important to have accurate knowledge of not only the primary proton interactions in the target, but also of the re-interactions of secondary particles.

NA61/SHINE has previously measured hadron production in interactions of 31~\GeVc protons with a thin carbon target for the benefit of the T2K experiment~\cite{NA612007Pi,NA612007K,NA612007Neutrals,na61_t2k_thin}. The NA61/SHINE experiment is also well suited to making measurements of the beam line interactions that dominate the neutrino production in the Fermilab long-baseline accelerator neutrino program, including the existing NuMI beam~\cite{numi_tdr}, which is initiated by 120 \GeVc primary protons, and the proposed Long-Baseline Neutrino Facility (LBNF) beam line~\cite{dune_physics} that will supply neutrinos for the Deep Underground Neutrino Experiment (DUNE)~\cite{dune_idr}, which will use 60-120 \GeVc primary protons. The current optimized beam line design for LBNF features a $\sim$2.2 m-long graphite target~\cite{ichep2018}, but beryllium and hybrid targets have been considered as well.

In DUNE, near the oscillation peak at a neutrino energy of 3 GeV, roughly half of the neutrinos are produced from the decays of secondary particles generated in the interactions of primary protons ($p \rightarrow X \rightarrow \nu$)~\cite{nuint2017}. The other half come from the decays of particles generated by the re-interactions of protons or hadrons (eg. $p \rightarrow X \rightarrow Y \rightarrow \nu$ ). For the LBNF optimized beam, each neutrino in the near detector results from an average of 1.8 interactions in the beam line (including the interaction of the primary proton)~\cite{fields_beyond2020}. After protons, the largest source of these interactions is pions with an average of 0.2 pion interactions contributing to each neutrino, and these pions typically have momenta in the range from roughly 10 \GeVc to 70 \GeVc .  

The current estimates of the flux uncertainties in DUNE~\cite{fields_beyond2020} near the oscillation maximum are dominated by uncertainties on existing $p+C$ measurements such as those described in Ref.~\cite{na49_pC}, proton and neutron interactions that are not covered by existing data and uncertainties on the re-interactions of pions and kaons. NA61/SHINE seeks to improve on these uncertainties by making improved measurements of proton interactions with neutrino target materials (with more phase space coverage and larger statistics) and by making measurements of meson interactions with target and beam line materials. With the exception of the HARP measurements~\cite{harppic12}, there is little existing data on the particle production spectra from interactions of mesons in the incident momentum range of interest for long-baseline neutrino experiments. This paper presents new results on the yields of particles resulting from the interactions of 60\,\GeVc  $\pi^{+}$ on carbon and beryllium targets recorded in 2016. 

Three types of results are presented in this paper. Section~\ref{sec:xsec} presents measurements of the integrated production and inelastic cross sections for \piC and \piBe interactions, and describes the uncertainties on these measurements. Section~\ref{sec:V0} describes measurements of the differential multiplicity of neutral hadrons (\kos, \lam and \alam) produced in these interactions, in bins of the momentum and angle of the produced hadron. Section~\ref{sec:dEdx} describes measurements of the differential multiplicity of the charged hadrons (\pip, \pim, \kp, \km and $p$) in bins of the momentum and angle of the produced hadron. Section~\ref{sec:spectrauncert} describes the systematic uncertainties on the results presented in Sections~\ref{sec:V0} and~\ref{sec:dEdx}.

\section{Detector Setup}
%Alysia

\begin{figure*}[t]
  \centering
  \includegraphics[width=\textwidth]{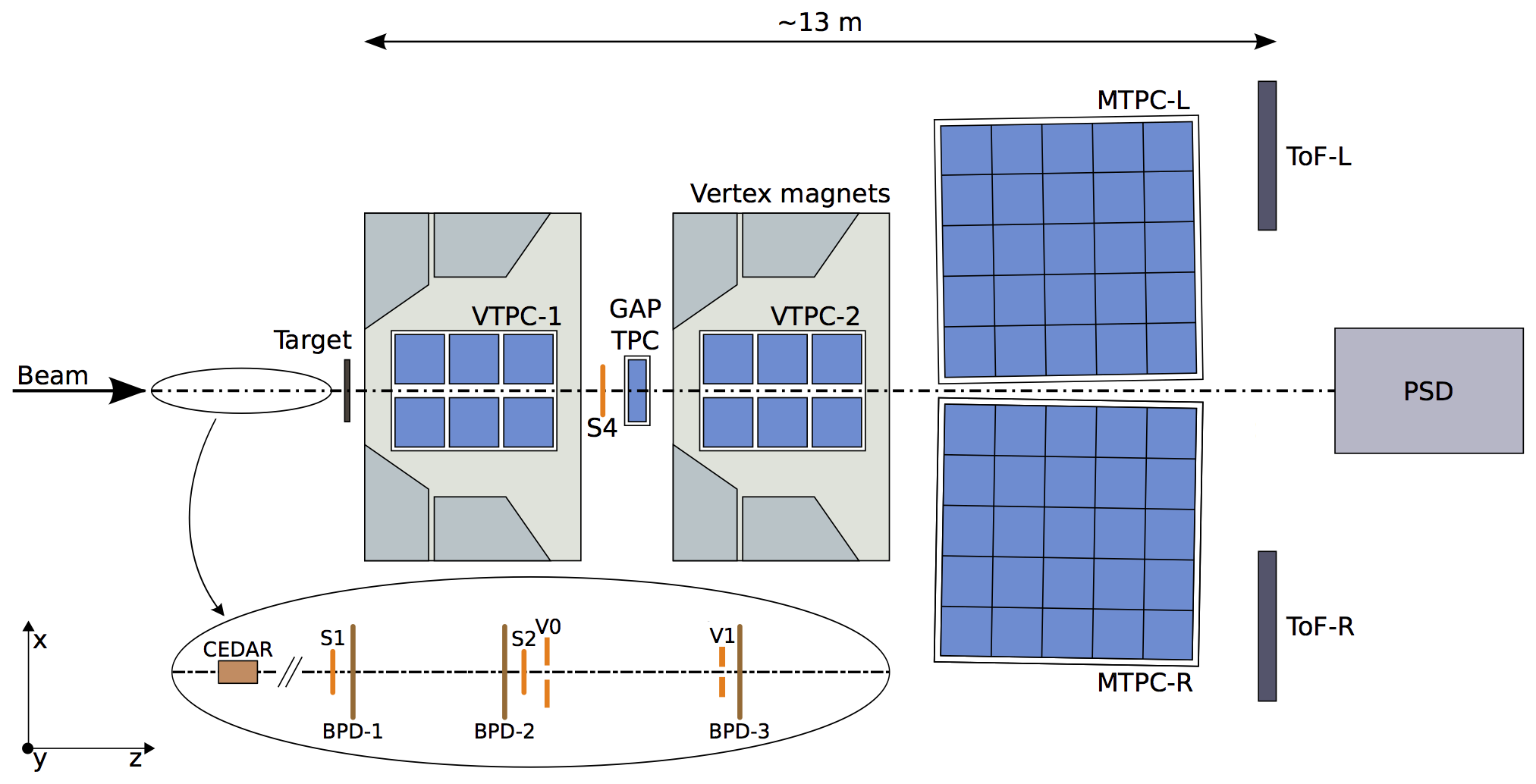}
 \caption{The schematic top-view layout of the NA61/SHINE experiment in the configuration used during the 2016 data taking. }\label{fig:exp_setup}
\end{figure*}

Located on a secondary beam line of CERN's Super Proton Synchrotron (SPS), NA61/SHINE probes the interactions of protons, pions, kaons and heavy ions with fixed targets. The 400\,\GeVc primary protons from the SPS beam strike a target 535\,m upstream of NA61/SHINE, generating the secondary beam. A system of magnets selects the desired beam momentum. Unwanted positrons and electrons are absorbed by a 4\,mm lead absorber.  
  
The NA61/SHINE detector~\cite{na61detector} is shown in Figure~\ref{fig:exp_setup}. In the 2016 operation configuration, the detector comprises four large Time Projection Chambers (TPCs) and a Time of Flight (ToF) system allowing NA61/SHINE to make spectral measurements of produced hadrons. Two of the TPCs, Vertex TPC 1 (VTPC-1) and Vertex TPC 2 (VTPC-2), are located inside superconducting magnets, capable of generating a combined maximum bending power of 9 T$\cdot$m. Downstream of the VTPCs are the Main TPC Left (MTPC-L) and Main TPC Right (MTPC-R).  Additionally, a smaller TPC, the Gap TPC (GTPC), is positioned along the beam axis between the two VTPCs. Two side time-of-flight walls, ToF-Left and ToF-Right, walls were present. Notably, the previously used ToF-Forward wall was not installed during the 2016 operation. The Projectile Spectator Detector (PSD), a forward hadron calorimeter, sits downstream of the ToF system.  

 The NA61/SHINE trigger system uses two scintillator counters (S1 and S2) to trigger on beam particles. The S1 counter provides the start time for all counters. Two veto scintillation counters ($V0$ and $V1$), each with a hole aligned to the beam, are used to remove divergent beam particles upstream of the target. The S4 scintillator with a 1 cm radius (corresponding to a particle scattering off of the target at an angle of 2.7 mrad) sits downstream of the target and is used to determine whether or not an interaction has occurred. A Cherenkov Differential Counter with Achromatic Ring Focus (CEDAR)~\cite{cedar, cedar82} identifies beam particles of the desired species. The CEDAR focuses the Cherenkov ring from a beam particle onto a ring of 8 Photomultiplier Tubes (PMTs). The pressure is set to a fixed value so that only particles of the desired species will trigger the PMTs, and typically, a coincidence of at least 6 PMTs is required to tag a particle for the trigger.  
 
The beam particles are selected by defining the beam trigger ($T_\mathrm{beam}$) as the coincidence of $S1\wedge S2\wedge\overline{V0}\wedge\overline{V1}\wedge CEDAR$.
The interaction trigger ($T_\mathrm{int}$) is defined by the coincidence of $T_\mathrm{beam}\wedge\overline{S4}$ to select beam particles which have interacted with the target. A correction factor will be discussed in detail in Section~\ref{sec:s4xsec} to correct for interactions that result in an S4 hit. Three Beam Position Detectors (BPDs), which are  proportional wire chambers, are located 30.39 m, 9.09 m and 0.89 m upstream of the target and determine the location of the incident beam particle to an accuracy of $\sim$100\,$\mu$m.

Interactions of $\pi^{+}$ beams were measured on thin carbon and beryllium targets. The carbon target was composed of graphite of density $\rho = 1.80\, \mbox{g/cm}^{3}$ with dimensions of 25\, mm (W) x 25\, mm (H) x 14.8\, mm (L), corresponding to roughly 3.1\% of a proton-nuclear interaction length. The beryllium target had a density of $\rho = 1.85\, \mbox{g/cm}^{3}$ with dimensions of 25\, mm (W) x 25\, mm (H) x 14.9\, mm (L), corresponding to roughly 3.5\% of a proton-nuclear interaction length. The uncertainties in the densities of the targets were found to be 0.69\% for the carbon target and 0.19\% for the beryllium target.

\section{Event Selection}

Several cuts were applied to events to ensure the purity of the samples and to control the systematic effects caused by beam divergence. The same event cuts are used for the integrated cross section and differential cross section analyses in order to ensure that the normalization constants obtained from the integrated cross section analysis are valid for calculating multiplicities in the differential cross section analyses. First, the so-called WFA (Wave Form Analyzer) cut was used to remove events in which multiple beam particles pass through the beam line in a small time frame. The WFA determines the timing of beam particles that pass through the S1 scintillator. If another beam particle passes through the beam line close in time to the triggered beam particle, it could cause a false trigger in the S4 scintillator and off-time tracks being reconstructed to the main interaction vertex. To mitigate these effects, a WFA cut of $\pm$ 2 $\mu$s is used.

\begin{figure*}[htbp]
\begin{center}
\includegraphics[width=0.45\textwidth]{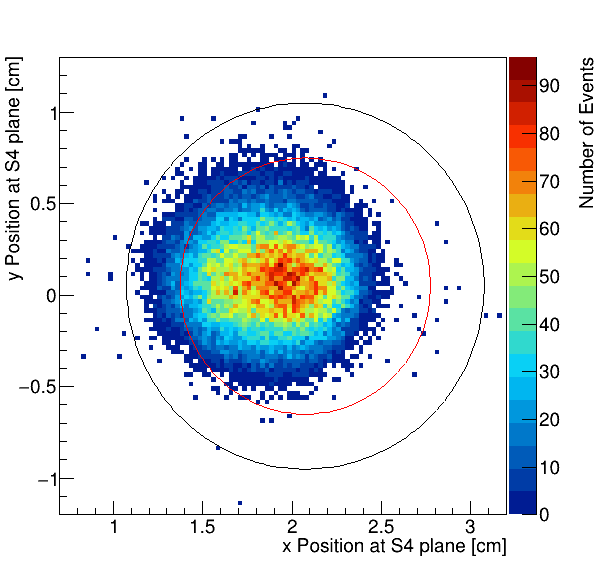}
\includegraphics[width=0.45\textwidth]{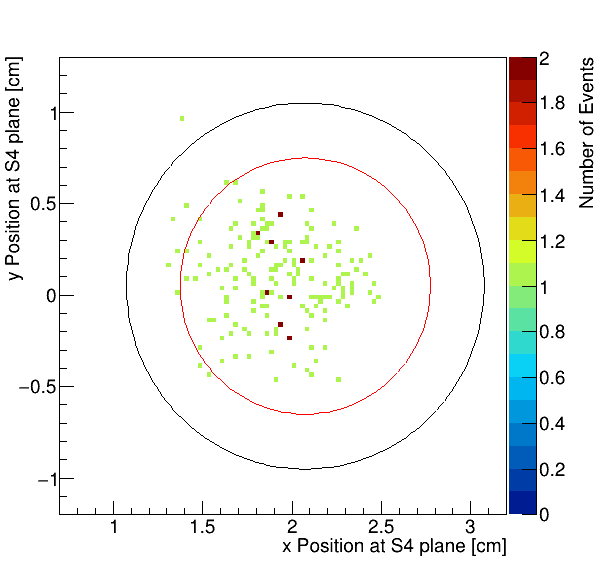}
\caption[Beam-on-S4 distributions for the 60~\GeVc beam]{Positions of BPD tracks extrapolated to the S4 plane in target-removed data runs from the \piC data set. 
The measured S4 position is shown as a black circle and the BPD radius cut is shown as a red circle in both figures.
\textit{Left}: Events taken by the beam trigger.
\textit{Right}: Events taken by the interaction trigger.}
\label{fig:bpdExtrap60_2016}
\end{center}
\end{figure*}

The measurements from the BPDs are important for estimating the effects of beam divergence on the integrated cross section measurements. To mitigate these effects, tracks are fitted to the reconstructed BPD clusters, and these tracks are extrapolated to the S4 plane. The so-called ``Good BPD" cut requires that each event includes a cluster in the most downstream BPD and that a track was successfully fit to the BPDs. Figure~\ref{fig:bpdExtrap60_2016} shows the resulting BPD extrapolation to the S4 plane for the 60~\GeVc $\pi^+$ beam. A radial cut was applied to the BPD tracks extrapolated to the S4, indicated by the red circles on Figure~\ref{fig:bpdExtrap60_2016}, in order to ensure that non-interacting beam particles strike the S4 counter. This corresponds to a trajectory within 0.7 cm of the S4 center (compared to the S4 radius of 1 cm). It can be seen from these distributions that the beam, veto counters and the S4 were well-aligned during the data taking.  

To begin the event selection, only unbiased $T_\mathrm{beam}$ events are considered for the integrated cross section analysis. For the integrated cross section analysis of the \piC (\piBe) data set, 191,099 (116,944) target inserted and 86,022 (58,551) target removed events were analyzed after the described selection. For the analysis of spectra, only $T_\mathrm{int}$ events are considered. For the spectra analysis of the \piC (\piBe) data set, 1,496,524 (1,096,003) target inserted and 86,764 (57,045) target removed events were selected.

\section{Integrated Inelastic and Production Cross Section Analysis}
\label{sec:xsec}

The total integrated cross section of hadron+nucleus interactions, $\sigma_\mathrm{tot}$, can be defined as the sum of the inelastic cross section, $\sigma_\mathrm{inel}$, and the coherent elastic cross section, $\sigma_\mathrm{el}$:
\begin{eqnarray}
\sigma_\mathrm{tot} = \sigma_\mathrm{inel} + \sigma_\mathrm{el}.\label{eq:tot_xsec}
\end{eqnarray} 
Coherent elastic scattering leaves the nucleus intact. The sum of all other processes due to strong interactions makes up the inelastic cross section. The inelastic cross section can be divided into the production cross section, $\sigma_\mathrm{prod}$, and the quasi-elastic cross section, $\sigma_\mathrm{qe}$:
\begin{eqnarray}
\sigma_\mathrm{inel} = \sigma_\mathrm{prod} + \sigma_\mathrm{qe}.\label{eq:prod_xsec}
\end{eqnarray} 
In this paper, production interactions are defined as processes in which new hadrons are produced. Quasi-elastic interactions include processes other than coherent elastic interactions in which no new hadrons are produced, mainly fragmentation of the nucleus. In this paper, measurements of the production cross section, $\sigma_\mathrm{prod}$, and inelastic cross section, $\sigma_\mathrm{inel}$, are presented for \piC and \piBe interactions. These cross section measurements are important for accelerator-based neutrino experiments and are needed to normalize the differential cross section yields that will be discussed in Sections~\ref{sec:V0} and~\ref{sec:dEdx}. This analysis closely follows the method described in Ref.~\cite{na61_prod_cross}, but with some differences, which will be discussed below. 

\subsection{Trigger Cross Section}
\label{sec:s4xsec}

For sufficiently thin targets, the probability $P$ of a beam particle interacting is approximately proportional to the thickness, $L$, of the target, the number density of the target nuclei, $n$, and the interaction cross section, $\sigma$:
\begin{eqnarray}\label{eq:p}
P = \frac{\text{ Number of interactions}}{\text{Number of incident particles}} = n\cdot L\cdot \sigma.
\end{eqnarray}
The density of nuclei can be written in terms of Avogadro's number, $N_{A}$, the material density, $\rho$, and the atomic mass, $m_{a}$:
\begin{equation}
n = \frac{\rho N_{A}}{m_{a}}.
\end{equation}

\begin{figure*}[htbp]
\begin{center}
\includegraphics[width=0.95\textwidth]{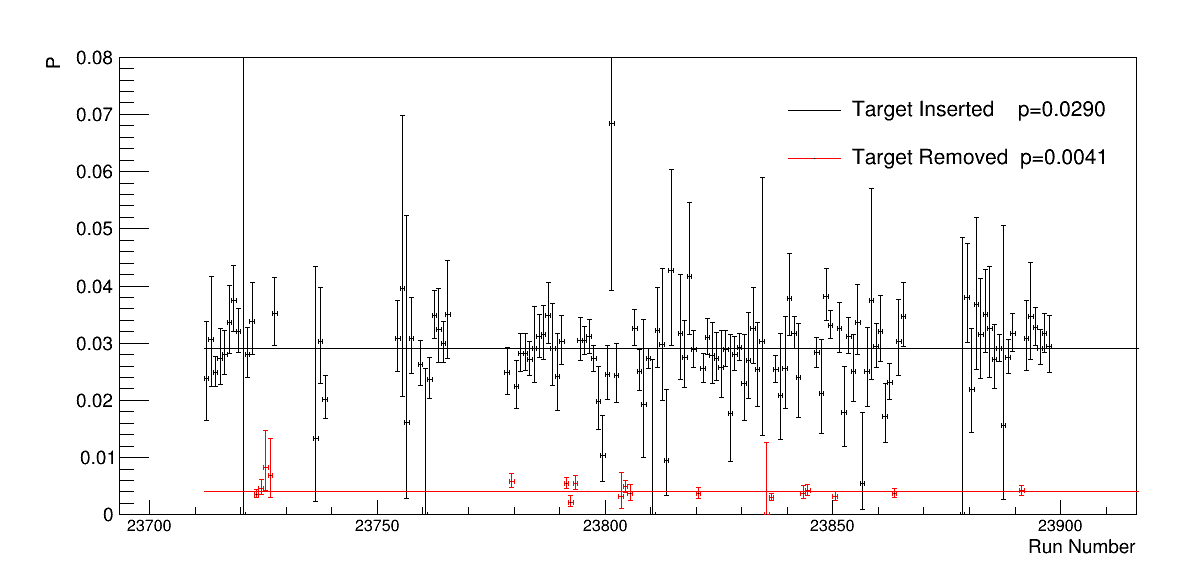}
\caption[Trigger interaction probabilities for \piC data]{Trigger interaction probabilities for the \piC data set for target-inserted and target-removed runs.}
\label{fig:ptint}
\end{center}
\end{figure*}

The counts of beam ($T_\mathrm{beam}$) and interaction triggers ($T_\mathrm{int}$) that pass the event selection can be used to estimate the trigger probability with the target inserted (\emph{I}) and with the target removed (\emph{R}):
\begin{eqnarray}
 P_\mathrm{T}^\mathrm{I,R} = \frac{N(T_\mathrm{beam} \land T_\mathrm{int})^\mathrm{I,R}}{N(T_\mathrm{beam})^\mathrm{I,R}} \label{eq:P_Tint}.
\end{eqnarray}
Figure~\ref{fig:ptint} shows an example of the trigger probabilities for each run for the \piC data set. The target-removed runs were interspersed throughout the target-inserted data runs to ensure they represented comparable beam conditions. The trigger rates show consistency over the course of the runs, which were recorded over a period of about three days. Table~\ref{tab:ptint} gives the trigger probabilities for both the target-inserted and target-removed samples of the \piC and \piBe data sets. 

\begin{table*}[htbp]
\centering
\begin{tabular}{cccccccc}
Interaction  & $p\,(\GeVc) $ & $P_\mathrm{Tint}^\mathrm{I}$ (\%) & $P_\mathrm{Tint}^\mathrm{R}$ (\%)\\
\hline
$ \pi^{+} + \mbox{C}$ & 60 & 2.90 $\pm$ 0.04 & 0.41 $\pm$ 0.02
\\%[0.2ex]
$ \pi^{+} + \mbox{Be}$& 60 & 3.28 $\pm$ 0.05 & 0.47 $\pm$ 0.03 \\%[0.2ex]
\end{tabular}
\caption[Trigger interaction probabilities]{This table presents the observed trigger interaction probabilities for both the target-inserted and target-removed samples of the \piC and \piBe data sets.} 
\label{tab:ptint}
\end{table*}

Taking into account the trigger probabilities with the target inserted and the target removed, $P_\mathrm{T}^\mathrm{I}$ and $P_\mathrm{T}^\mathrm{R}$, the corrected trigger probability, $P_\mathrm{trig}$, can be obtained:
\begin{eqnarray}
 P_\mathrm{trig} = \frac{P_\mathrm{T}^\mathrm{I} - P_\mathrm{T}^\mathrm{R}}{1 - P_\mathrm{T}^\mathrm{R}}. \label{eq:Pint}
\end{eqnarray}
Analogous to Equation~\ref{eq:p}, the trigger cross section $\sigma_\mathrm{trig}$ is defined as: 
\begin{eqnarray}
 \sigma_\mathrm{trig} = \frac{m_{a}}{\rho L_\mathrm{eff} N_\mathrm{A}}\cdot P_\mathrm{trig}\label{eq:sigma},
\end{eqnarray}
where  the beam attenuation is taken into account by replacing $L$ with $L_\mathrm{eff}$. The effective target length can be calculated using the absorption length, $\lambda_\mathrm{abs}$:
\begin{eqnarray}
L_\mathrm{eff} = \lambda_\mathrm{abs}(1 - e^{-L/\lambda_\mathrm{abs}})\label{eq:Leff},
\end{eqnarray}
where
\begin{eqnarray}
\lambda_\mathrm{abs} = m_{a}/(\rho N_\mathrm{A} \sigma_\mathrm{trig}).\label{eq:lam}
\end{eqnarray}
By combining Equations~\ref{eq:sigma}, ~\ref{eq:Leff} and ~\ref{eq:lam},  $\sigma_\mathrm{trig}$ can be rewritten as
\begin{eqnarray}
 \sigma_\mathrm{trig} = \frac{m_{a}}{\rho L N_\mathrm{A}} \text{ln}(\dfrac{1}{1-P_\mathrm{trig}}).\label{eq:sigma1}
\end{eqnarray}

\subsection{S4 Correction Factors}

The trigger cross section takes into account the interactions where the resulting particles miss the S4 scintillator.  But even when there has been a production or quasi-elastic interaction in the target, there is a possibility that a forward-going particle will strike the S4 counter. Moreover, not all elastically scattered beam particles strike the S4. The trigger cross section must be corrected to account for these effects. Combining Equations~\ref{eq:tot_xsec} and~\ref{eq:prod_xsec}, the trigger cross section can be related to the production cross section through Monte Carlo (MC) correction factors as follows:
\begin{equation}
\sigma_\mathrm{trig}=\sigma_\mathrm{prod}\cdot f_\mathrm{prod}+\sigma_\mathrm{qe}\cdot f_\mathrm{qe}+\sigma_\mathrm{el}\cdot f_\mathrm{el}\ ,\label{eq:s4_fact}
\end{equation}
where $f_\mathrm{prod}$, $f_\mathrm{qe}$ and $f_\mathrm{el}$ are the fractions of production, quasi-elastic and elastic events that miss the S4 counter. The cross sections $\sigma_\mathrm{qe}$ and $\sigma_\mathrm{el}$ are also estimated from MC.  
Equation~\ref{eq:s4_fact} can be rewritten to obtain $\sigma_\mathrm{prod}$ and $\sigma_\mathrm{inel}$ as:
\begin{equation}
\sigma_\mathrm{prod}= \frac{1}{f_\mathrm{prod}}( \sigma_\mathrm{trig}  - \sigma_\mathrm{qe}\cdot f_\mathrm{qe} - \sigma_\mathrm{el}\cdot f_\mathrm{el})\label{eq:sig_prod}
\end{equation}
and
\begin{equation}
%\sigma_{inel}= \frac{1}{f_{inel}}( \sigma_{trig}  -  \sigma_{el}\cdot f_{el}).\label{eq:sig_inel}
\sigma_\mathrm{inel}= \frac{1}{f_\mathrm{inel}}( \sigma_\mathrm{trig}  -  \sigma_\mathrm{el}\cdot f_\mathrm{el}).\label{eq:sig_inel}
\end{equation}

A GEANT4  detector simulation~\cite{Agostinelli:2002hh, Allison:2006ve, Allison:2016lfl} using GEANT4 version 10.4 with physics list FTFP\_BERT was used to estimate the MC correction factors discussed above. The MC correction factors obtained for \piC and \piBe interactions are presented in Table~\ref{tab:MC_factors}. 

\begin{table*}[htbp]
\centering
\begin{tabular}{cccccccc}
Interaction  & $p$ & \multicolumn{6}{c}{Monte Carlo Correction Factors}\\
\cline{3-8}
                    & (\GeVc) & $\sigma_\mathrm{el}$ (mb)& $f_\mathrm{el}$ & $\sigma_\mathrm{qe}$ (mb) & $f_\mathrm{qe}$ & $f_\mathrm{prod}$ & $f_\mathrm{inel}$\\ 
\hline
$ \pi^{+} + \mbox{C}$ &  60 & 54.1 & 0.268 & 15.9 & 0.813 & 0.976 & 0.961\\%[0.2ex]
$ \pi^{+} + \mbox{Be}$& 60 & 39.6 & 0.229 & 13.7 & 0.813 & 0.975 & 0.960\\%[0.2ex]
\end{tabular}
\caption{Monte Carlo correction factors obtained for analyzing \piC and \piBe interactions.
} 
\label{tab:MC_factors}
\end{table*}

\subsection{Beam Composition}
\label{sec:beamComp2016}

For the analyses of \piC and \piBe interactions recorded in 2016, the beam composition could be constrained better than in the analysis of interactions recorded in 2015 by NA61/SHINE as discussed in~\cite{na61_prod_cross}. Simulations of the H2 beam line show that the population of muons in the 60~\GeVc secondary hadron beam used to record these interactions is at the level of 1.5$\pm$0.5\%~\cite{nikos}. Nearly all of the muons come from decays of 60~\GeVc pions, so they have a minimum energy of 34~\GeVc. GEANT4 simulations were run to estimate the target-inserted and target-removed trigger rates due to muons, $P^{I}_{\mu}$ and $P^{R}_{\mu}$. These simulations took the momentum distribution of muons into account. Additional H2 beam line simulations were run to more precisely estimate the level of positron contamination in the beam~\cite{Nikos2}. A conservative estimate of $0.5\% \pm 0.5\%$ was attributed to this contamination. The trigger rates due to positrons, $P^{I}_{e}$ and $P^{R}_{e}$, were also estimated with GEANT4 simulations. The effect of muon and positron contamination on the trigger cross section was estimated as follows:
\begin{equation}
P_\mathrm{T}^\mathrm{\pi^{+}} = (P_\mathrm{T}-P_{e}\cdot f_{e} - P_{\mu}\cdot f_{\mu})/f_{\pi}  \quad \mathrm{(Target\ I, R)} \ ,
\end{equation}
where $f_e = 0.005$, $f_{\mu} = 0.015$ and $f_{\pi} = 0.98$. The resulting corrections applied to $\sigma_\mathrm{prod}$ ($\sigma_\mathrm{inel}$) were +0.3\% (+0.3\%) for \piC and +1.1\% (1.0\%) for \piBe. 

\subsection{Systematic Uncertainties}

The integrated cross section results were evaluated for a number of possible systematic effects. The sources of uncertainty having a non-negligible effect on the results are the uncertainty in the density of the target, the uncertainty in the S4 size, the uncertainty on the beam composition and uncertainties on the S4 correction factors. The procedures used to evaluate these sources of systematic uncertainties were discussed in~\cite{na61_prod_cross}, so they will not be discussed here.

\subsubsection{Breakdowns of the Integrated Cross Section Uncertainties}

The target density uncertainties, S4 size uncertainties, beam composition uncertainties and S4 correction factor uncertainties associated with the production and inelastic cross sections measurements for \piC and \piBe interactions are presented in Tables \ref{tab:sys_prod} and \ref{tab:sys_inel}. 

\begin{table*}[htbp]
\centering
\begin{tabular}{cccccccc}
  &  & \multicolumn{4}{c}{Systematic uncertainties for $\sigma_\mathrm{prod}$ (mb)}& & \\
\cline{3-7}
                    &       $p$      &  & S4  & Beam  & MC  & Total Syst. & Model\\ 
    Interaction  & (\GeVc) & Density & Size  &Purity & Stat. & Uncer. & Uncer.\\ 

\hline
$ \pi^{+} + \mbox{C}$&  60 & $\pm 1.3$ & $ \pm ^{1.1} _{1.2}$ & $\pm ^{1.5} _{1.5}$ & $\pm 0.2$ & $\pm ^{2.3}_{2.4}$ & $\pm ^{0.2} _{3.8}$\\[0.2ex]
$ \pi^{+} + \mbox{Be}$& 60 & $\pm 0.3$ & $\pm ^{0.8}_{0.9}$ & $\pm^{0.7}_{0.7}$  & $\pm 0.1$ & $\pm  ^{1.2} _{1.2}$ & $\pm ^{0.1}_{3.5}$\\[0.2ex]
\end{tabular}
\caption[Production cross section uncertainties breakdown]{Breakdown of systematic uncertainties for the production cross section measurements of \piC and \piBe interactions. 
} 
\label{tab:sys_prod}
\end{table*}

\begin{table*}[htbp]
\centering
\begin{tabular}{cccccccc}
       &  & \multicolumn{4}{c}{Systematic uncertainties for $\sigma_\mathrm{inel}$ (mb)}& & \\
\cline{3-7}
                    &       $p$      & & S4  & Beam  & MC  & Total Syst. & Model\\ 
    Interaction  & (\GeVc) & Density & Size  &Purity & Stat. & Uncer. & Uncer.\\ 
\hline
$ \pi^{+} + \mbox{C}$ & 60 & $\pm 1.4$ & $\pm^{1.1} _{1.2}$ & $\pm^{1.6} _{1.6}$& $\pm 0.2$   & $\pm^{2.4} _{2.4}$ & $\pm^{0.2} _{2.8}$\\[0.2ex]
$ \pi^{+} + \mbox{Be}$& 60 & $\pm 0.3 $& $\pm^{0.9} _{0.9}$ & $\pm^{0.7} _{0.7}$& $\pm 0.1$   & $\pm^{1.2} _{1.2}$ & $\pm^{0.1} _{2.5}$\\[0.2ex]
\end{tabular}
\caption[Inelastic cross section uncertainties breakdown]{Breakdown of systematic uncertainties for the inelastic cross section measurements of \piC and \piBe interactions. 
} 
\label{tab:sys_inel}
\end{table*}

\subsection{Integrated Cross Section Results}

\begin{figure*}[htbp]
\begin{center}
\includegraphics[width=0.85\textwidth]{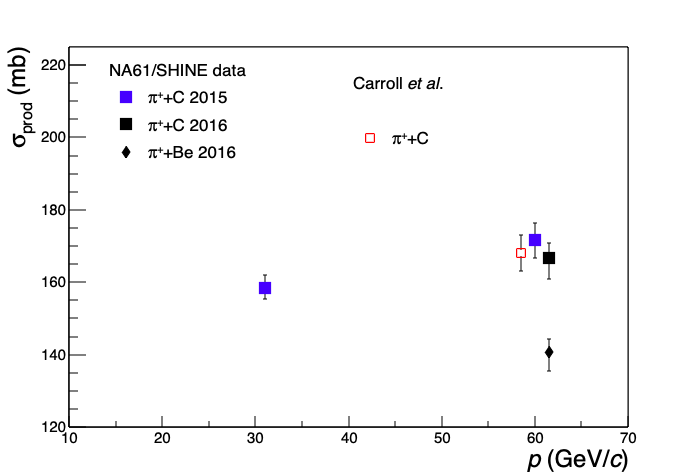}
\caption[Production cross section measurements summary]{Summary of production cross section measurements.
The results are compared to previous results from NA61/SHINE~\cite{na61_prod_cross} and Carroll \textit{et al}.~\cite{Carroll}.
}
\label{fig:ProdXsec}
\end{center}
\end{figure*}

Measurements of production cross sections for \piC and \piBe are summarized in Table~\ref{t:nuprodcross} along with statistical, systematic and physics model uncertainties. The production cross section of \piC interactions was found to be 166.7\,mb, and the production cross section of \piBe interactions was found to be 140.6\,mb. The result obtained for interactions of \piC with these 2016 data was lower compared to the result obtained with the 2015 data~\cite{na61_prod_cross}, but it is within the estimated uncertainty. Reasons for this difference could be due to the difference in the detector setup, the different target used and statistical fluctuations. These results, the results obtained by NA61/SHINE from data recorded in 2015 and the measurements of Carroll \textit{et al}.~\cite{Carroll} are compared in Figure~\ref{fig:ProdXsec}.

\begin{figure*}[htbp]
\begin{center}
\includegraphics[width=0.85\textwidth]{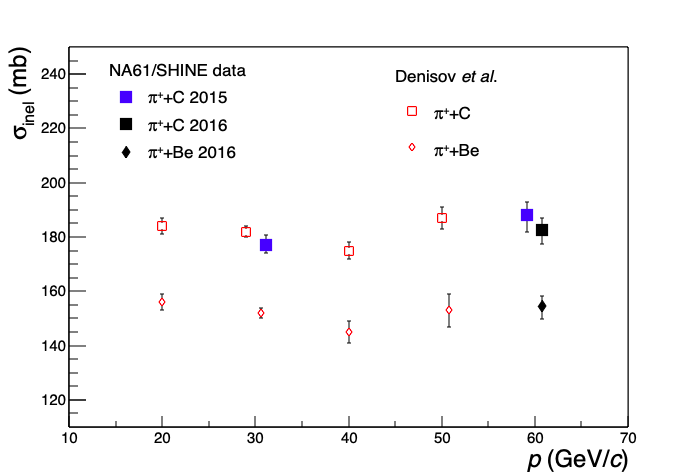}
\caption[Inelastic cross section measurements summary]{Summary of inelastic cross section measurements.
The results are compared to previous results from NA61/SHINE~\cite{na61_prod_cross} and  Denisov \textit{et al}.~\cite{Denisov:1973zv} .
}
\label{fig:InelXsec}
\end{center}
\end{figure*}

The measurements of inelastic cross sections for \piC and \piBe are summarized in Table~\ref{t:nuinelcross} along with statistical, systematic and physics model uncertainties. The inelastic cross section of \piC was found to be 182.7\,mb, and the inelastic cross section of \piBe was found to be 154.4\,mb. Again, the result obtained for interactions of \piC with these 2016 data was lower compared to the result obtained with the 2015 data~\cite{na61_prod_cross}, but it is within the estimated uncertainty. These results, the results obtained by NA61/SHINE from data recorded in 2015 and the measurements of Denisov \textit{et al}.~\cite{Denisov:1973zv} are compared in Figure~\ref{fig:InelXsec}.

\begin{table}[htbp]
\centering
\begin{tabular}{ccccccc}
Interaction  & $p$& \multicolumn{4}{c}{Production cross section (mb)} \\ 
                     & (\GeVc)  &$\sigma_\mathrm{prod}$ & $\Delta_\mathrm{stat}$& $\Delta_\mathrm{syst}$ & $\Delta_\mathrm{model}$ & $\Delta_\mathrm{total}$ \\
%\midrule
\hline
$ \pi^{+} + \mbox{C}$&  60 & 166.7 & $\pm 3.5$ & $\pm^{2.3} _{2.4}$ & $\pm^{0.2} _{3.9}$ &$\pm^{4.2} _{5.8}$ \\[0.2ex]
$ \pi^{+} + \mbox{Be}$& 60 & 140.6 & $\pm 3.5$ & $\pm^{1.2} _{1.2}$ & $\pm^{0.1} _{3.5}$ & $\pm^{3.7} _{5.1}$\\[0.2ex]
\end{tabular}
\caption[Production cross section measurements]{Production cross section measurements of \piC and \piBe interactions are presented. The central values as well as the statistical ($\Delta_\mathrm{stat}$), systematic ($\Delta_\textrm{syst}$) and model ($\Delta_\mathrm{model}$) uncertainties are shown.  The total uncertainties ($\Delta_\textrm{total}$) are the sum of the statistical, systematic and model uncertainties in quadrature.}
\label{t:nuprodcross}
\end{table}

\begin{table}[htbp]
\centering
\begin{tabular}{ccccccc}
Interaction  & $p$ & \multicolumn{4}{c}{Inelastic cross section (mb)} \\ %\cline{3-6}  
                     & (\GeVc)  &$\sigma_\mathrm{inel}$ & $\Delta_\mathrm{stat}$& $\Delta_\mathrm{syst}$ & $\Delta_\mathrm{model}$ & $\Delta_\mathrm{total}$ \\
\hline
$ \pi^{+} + \mbox{C} $ & 60 &  182.7 & $\pm 3.6$ & $\pm^{2.4} _{2.4}$ & $\pm^{0.2} _{2.8}$ & $\pm^{4.3} _{5.2}$ \\[0.2ex]
$ \pi^{+} + \mbox{Be}$ & 60 &  154.4 & $\pm 3.5$ & $\pm^{1.2} _{1.2}$ & $\pm^{0.1} _{2.5}$ & $\pm^{3.7} _{4.5}$ \\[0.2ex]
\end{tabular}
\caption[Inelastic cross section measurements]{Inelastic cross section measurements of \piC and \piBe interactions are presented. The central values as well as the statistical ($\Delta_\mathrm{stat}$), systematic ($\Delta_\textrm{syst}$) and model ($\Delta_\mathrm{model}$) uncertainties are shown. The total uncertainties ($\Delta_\textrm{total}$) are the sum of the statistical, systematic and model uncertainties in quadrature.}
\label{t:nuinelcross}
\end{table}

\section{Analysis of Neutral Hadron Spectra}
\label{sec:V0}

NA61/SHINE is able to identify a number of species of weakly-decaying neutral hadrons by tracking their charged decay products. The simplest decay topology NA61/SHINE can identify is the V$^{0}$ topology. This topology refers to track topologies in which an unobserved neutral particle decays into two child particles, one positively charged and one negatively charged, observed by the tracking system. This paper presents differential production cross section measurements of produced \kos, \lam and \alam in interactions of \piC and \piBe using a V$^{0}$ analysis. 

\subsection{Selection of V$^{0}$ Candidates}

To start with, every pair of one positively charged and one negatively charged track with a distance-of-closest approach less than 5 cm is considered as a V$^{0}$ candidate. Of course, many of these V$^{0}$ candidates are not true V$^{0}$s. For example, a V$^{0}$ candidate might consist of two tracks that come from the main interaction point, the child tracks might come from two different vertices or the child tracks might come from a parent track, which is not a neutral particle. Additionally, photons converting to $e^+e^-$ pairs make up part of the V$^{0}$ sample.

\subsubsection{Topological Cuts}

The topological cuts are designed to reduce the number of false V$^{0}$s in the collection of V$^{0}$ candidates and to remove V$^{0}$ candidates that have poorly fitted track variables. Only V$^{0}$ candidates that have a reconstructed V$^{0}$ vertex downstream of the target are considered.

The second topological selection is the requirement that both child tracks have at least 20 reconstructed TPC clusters and that at least 10 of those clusters belong to the VTPCs. This cut ensures that the reconstructed kinematics of the decay are reliable.

The third topological cut is the impact parameter cut, which removes many false V$^{0}$ candidates. This selection allows an impact parameter from between the extrapolated  V$^{0}$s track and the main interaction vertex of up to 4 cm in the x dimension and up to 2 cm in the y dimension.

\subsubsection{Purity Cuts}

The purity cuts are designed to separate the desired neutral hadron species from other neutral species, as well as to remove additional false V$^{0}$ candidates. The first two purity cuts are applied in the same way to \kos, \lam and \alam. This first selection requires the reconstructed z position of the V$^{0}$ vertex to be at least 3.5 cm downstream of the target center. This cut removes many of the V$^{0}$ candidates coming from the main interaction vertex and neutral species that decay more quickly than \kos, \lam or \alam.

Photons undergoing pair production ($\gamma \rightarrow e^+ e^-$) are present in the V$^{0}$ sample. Because the photon is massless, the transverse momentum of the decay is: 
\begin{equation}
p_{T} = |p_T^{+}| + |p_T^{-}| = 0 ~\GeVc.
\end{equation}
In order to remove most of these photons from the sample, the second purity cut requires a $p_{T} > 0.03 ~\GeVc$.

\subsubsection{Purity Cuts for the Selection of $K^0_S$}

At this point, it is necessary to assume a decay hypothesis. For \kos, the hypothesis is $\kos \rightarrow \pip \pim$. Therefore, it is assumed that the V$^{0}$ particle has a mass of $m_{\kos} = 0.498$ \GeVcc and the child particles have a mass of $m_{\pipm} = 0.140$ \GeVcc~\cite{PhysRevD.98.030001}. 

\begin{figure*}[htbp]
\begin{center}
\includegraphics[width=0.5\textwidth]{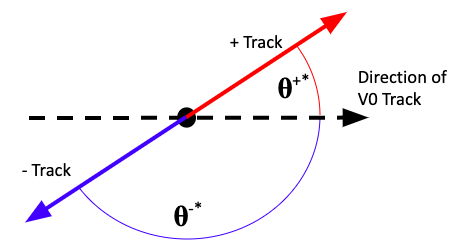}
\caption[Decay frame angles]{This cartoon shows the relevant angles in V$^{0}$ decays in the rest frame of the V$^{0}$. The child particles decay back to back in this frame. The angle at which the positively charged particle is emitted is $\theta^{+*}$, and the angle at which the negatively charged particle is emitted is $\theta^{-*}$.
}
\label{fig:DecayFrameAngles}
\end{center}
\end{figure*}

To remove \lam and \alam from the \kos sample, cuts on the angles that the child particle tracks make with the V$^{0}$ track in the decay frame are applied to the sample. These angles are represented in Figure~\ref{fig:DecayFrameAngles}. In order to remove \lam, $\cos{\theta^+}^* < 0.8$ is required and to remove \alam, $\cos{\theta^-}^* < 0.8$ is required.

The next selection is an allowed range of the invariant mass. The invariant mass is calculated with the reconstructed momenta, assumed masses and energies of the child particles:
\begin{equation}
m_{+-} = \sqrt{m_{+}^2+m_{-}^2+2 (E_+ E_- - \overrightarrow{p_+} \cdot \overrightarrow{p_-})}.
\end{equation}
The invariant mass range cut removes V$^{0}$ candidates with unreasonable values of $M_{\pip \pim}$, but is wide enough to allow a reliable fit to the background invariant mass distribution. For \kos, this range is chosen to be $[0.4, 0.65]$ \GeVcc. 

The final cut applied to the \kos selection is a cut on the proper decay length, $c\tau$. The proper decay length can be calculated with the estimated momentum of the V$^{0}$, $p$, the assumed mass, $m$, and the reconstructed length of the V$^{0}$ track, $L$:
\begin{equation}
c\tau = \frac{pL}{mc}.
\end{equation}
The purpose of this cut is to further reduce the number of false V$^{0}$s and more quickly decaying neutral species. The chosen cut is $c\tau > 0.67$ cm, which is a quarter of the proper decay length provided by the PDG~\cite{PhysRevD.98.030001}, 2.68 cm. 

\subsubsection{Purity Cuts for the Selection of \lam and \alam}

An invariant mass range cut and a proper decay length cut are used in the purity selection of \lam and \alam. The invariant mass hypothesis for the \lam decay is $\lam \rightarrow p \pi^-$ and the hypothesis for the \alam is $\alam \rightarrow \bar{p} \pi^+$. An invariant mass range of $[1.09, 1.215]$ \GeVcc is used in both the \lam and \alam analyses.

A proper decay length cut is also applied to the \lam and \alam selection. The chosen cut is $c\tau > 1.97$ cm, which is a quarter of the proper decay length given by the PDG~\cite{PhysRevD.98.030001}, 7.89 cm. 

\subsubsection{Armenteros-Podolansky Distributions}
\begin{figure*}[htbp]
\begin{center}
\includegraphics[width=0.45\textwidth]{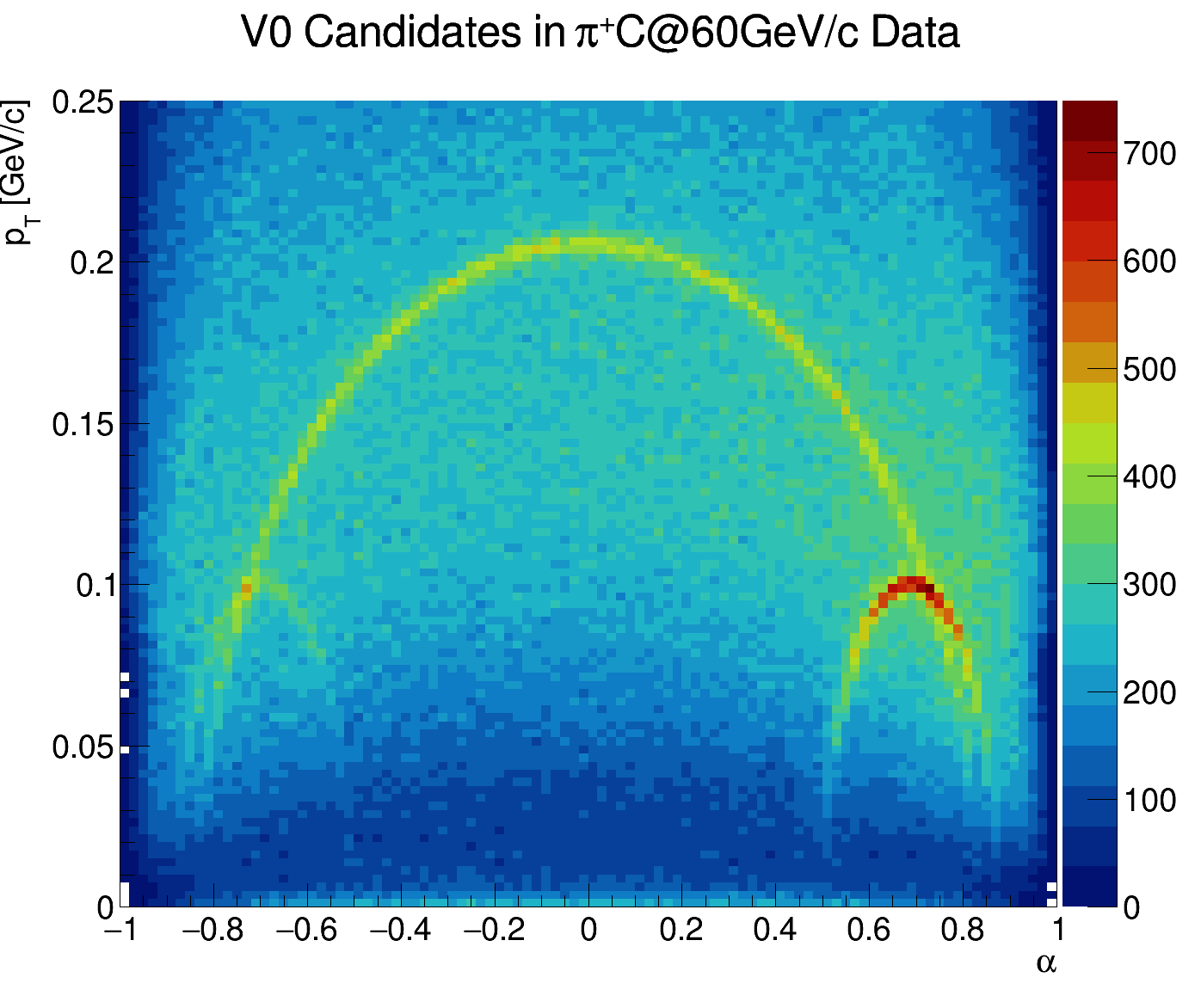}
\includegraphics[width=0.45\textwidth]{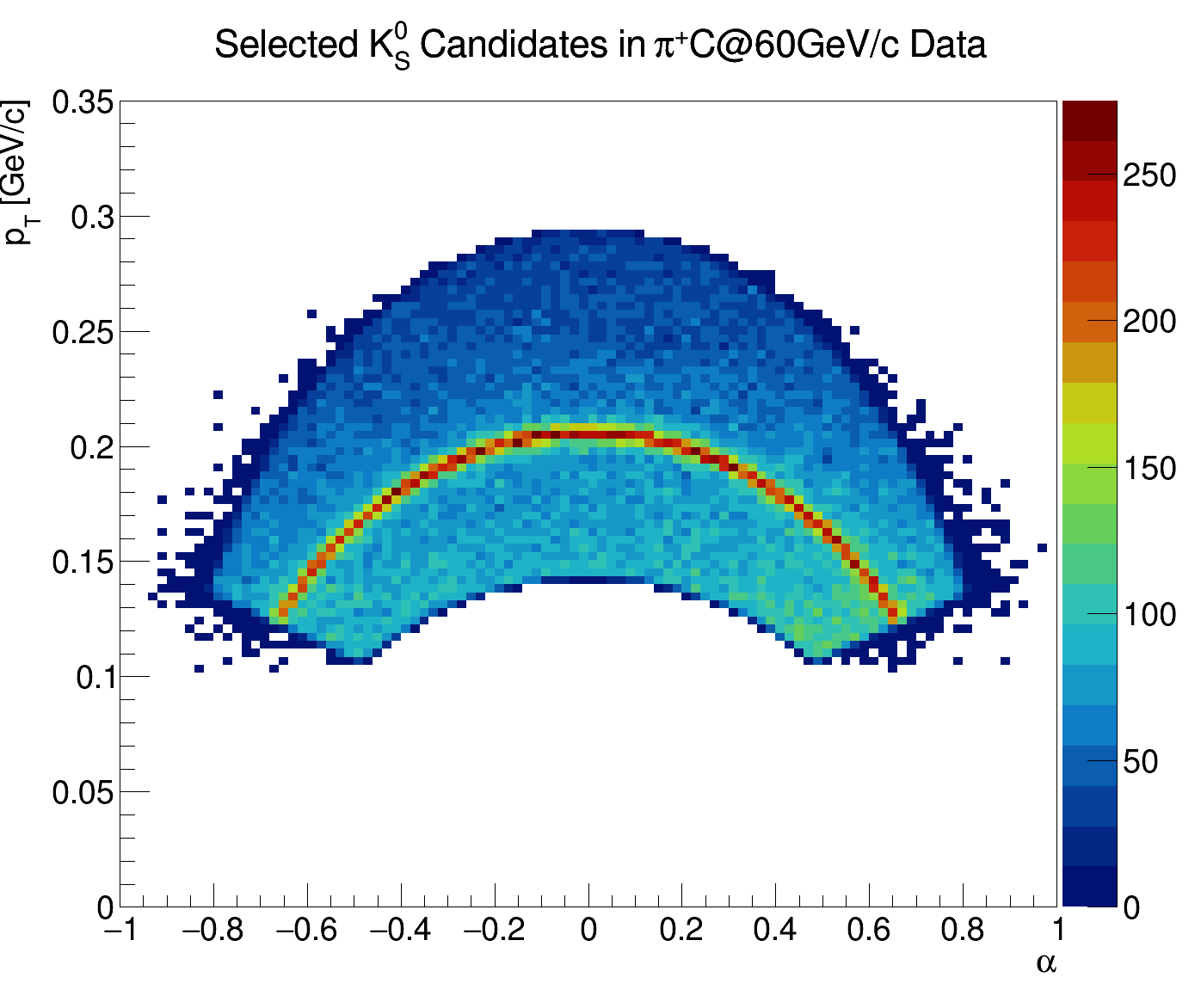}
\includegraphics[width=0.45\textwidth]{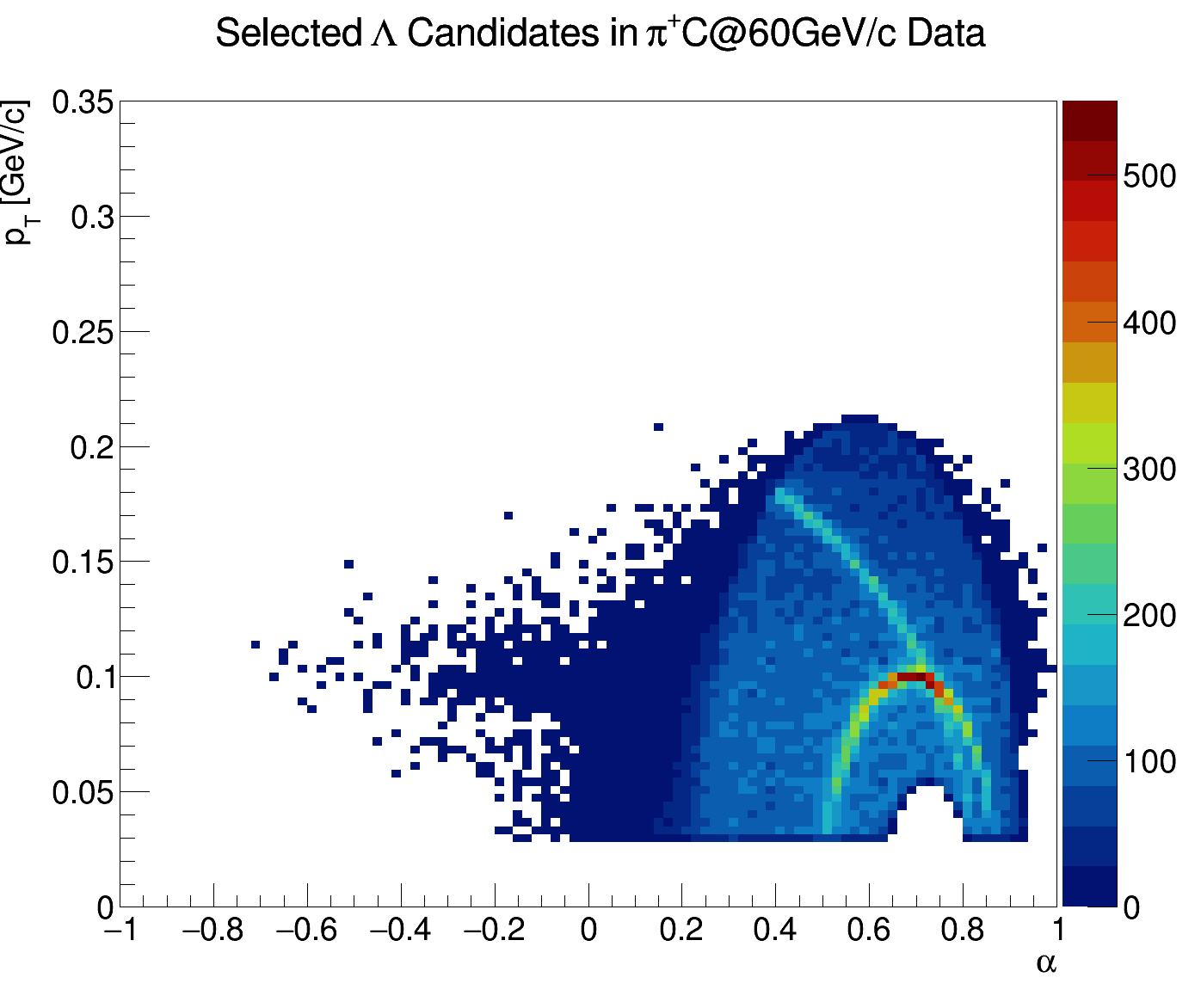}
\includegraphics[width=0.45\textwidth]{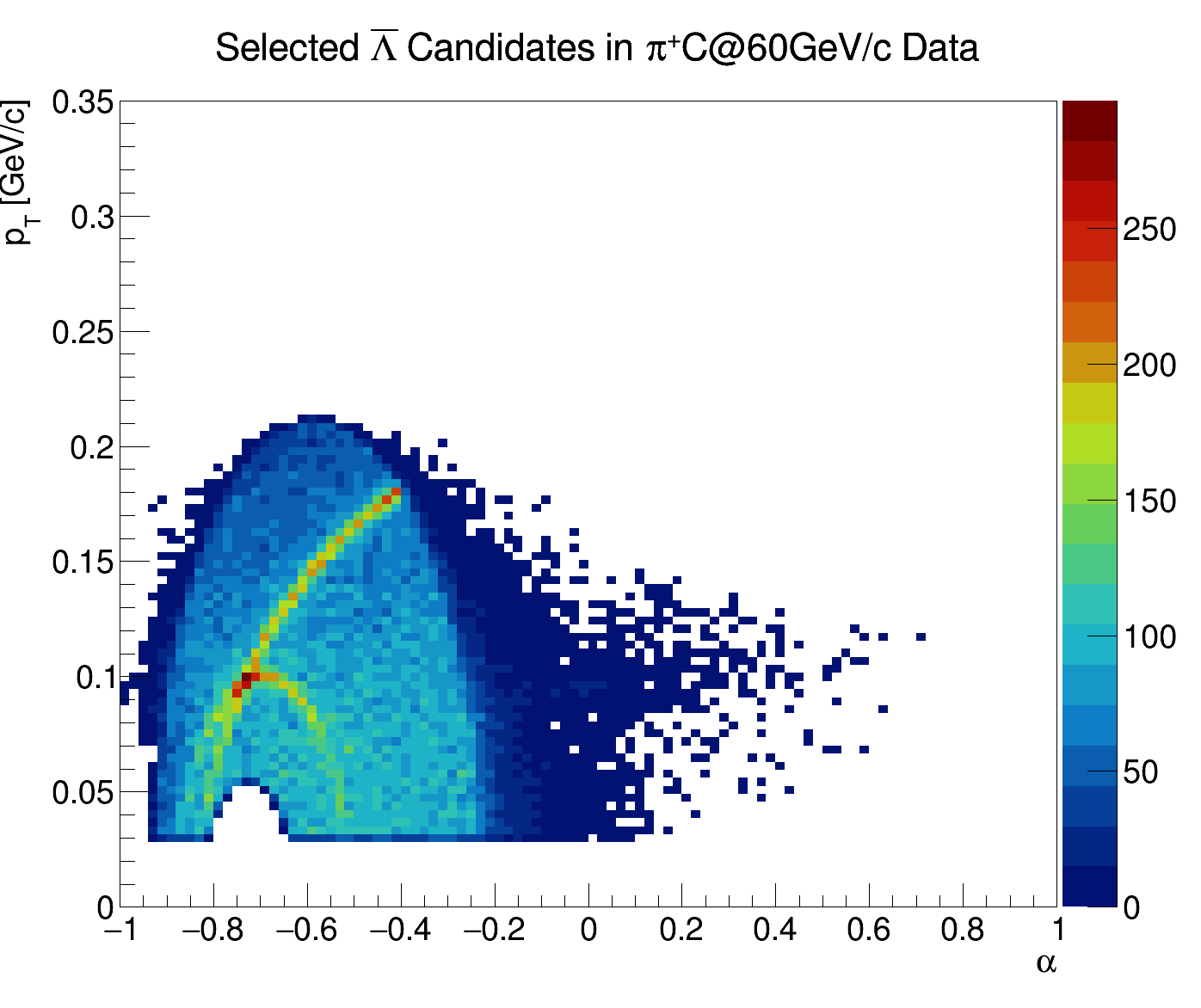}
\caption[Armenteros-Podolansky distributions]{The Armenteros-Podolanksy distribution of the V$^{0}$ candidates in the \piC analysis before selection cuts were applied is shown in the \textit{top left}. The distribution is shown after selection cuts are applied for the \kos analysis (\textit{top right}), \lam analysis (\textit{bottom left}) and \alam analysis (\textit{bottom right}).
}
\label{fig:PAPlot_PiC60}
\end{center}
\end{figure*}

The effect of these selections on the V$^{0}$ candidates can be visualized with Armenteros-Podolansky distributions, which are distributions of $\alpha$ vs. $p_{T}$. The parameter $\alpha$ is the asymmetry in the longitudinal momenta of the child tracks with respect to the V$^{0}$ track:
\begin{equation}
\alpha = \frac{p^+_L - p^-_L}{p^+_L + p^-_L}.
\end{equation}
Figure~\ref{fig:PAPlot_PiC60} shows the V$^{0}$ candidates coming from \piC interactions before the V$^{0}$ selection cuts were applied and after the selection cuts were applied for the \kos, \lam and \alam analyses. It can be seen that the \lam and \alam candidates include part of the \kos spectra. These \kos are separated out from \lam and \alam during the fitting procedure discussed in the following section.

\subsection{Fitting of Invariant Mass Distributions}

After applying the selection cuts for each particle species, the V$^{0}$ candidates are placed into the kinematic bins. For each of these kinematic bins, invariant mass distributions consist of both true \kos, \lam or \alam (signal) and the remaining background vertices. The objective of the fitting routine is to determine the number of true \kos, \lam and \alam in these invariant mass distributions. These fits are performed the same way on target-inserted and target-removed samples.

\subsubsection{Signal Model}
\label{sec:sModels}

In order to model the invariant mass distribution of \kos, \lam and \alam coming from the main interactions, template invariant mass distributions were derived from a GEANT4 MC production using the physics list FTFP\_BERT. V$^{0}$ vertices are reconstructed, selected and binned in the same way as was done with the data. For each kinematic bin, MC templates are formed from the distributions  of invariant mass from true \kos, \lam and \alam. These template distributions, $g_{MC}(m)$, are generated for both target-inserted and target-removed MC productions and were observed to peak at the known values of the \kos and \lam masses. In order to account for shifts in the invariant mass peaks and distortions of the signal shape due to misreconstruction of track variables and other possible effects, a mass shift, $m_0$, and a smearing are applied to $g_{MC}(m)$. The smearing is applied by convolving $g_{MC}(m)$ with a unit gaussian distribution with width $\sigma_s$. The parameters, $m_0$ and $\sigma_s$ are allowed to vary for each kinematic bin and were observed to be small compared to the widths of the invariant mass distributions. The full signal distribution can be written as: 
\begin{equation}
f_s(m; m_0, \sigma_s) = g_{MC}(m-m_0) \bigotimes \frac{1}{\sqrt{2\pi}\sigma_s}\exp{-\frac{(m-m_0)^2}{2\sigma_s^2}}.
\end{equation}

\subsubsection{Background Model}
\label{sec:bgModels}

It was observed that the shapes of the backgrounds in the invariant mass distributions vary among the \kos, \lam and \alam selection as well as among the kinematic bins. The background model was required to be flexible enough to account for the variation of background shapes in all of the kinematic bins for \kos, \lam and \alam. A second order polynomial was chosen to be used to fit the background distributions.

\subsubsection{Fitting Strategy}

In order to fit for the signal and background contributions to the invariant mass distributions, a continuous log-likelihood function is constructed:
\begin{equation}
\log{L} = \sum_{\mbox{V$^{0}$ Candidates}}{\log{F(m; \theta)}} ,
\end{equation}
where
\begin{equation}
F(m; \theta) = c_{s} f_{\text{s}}(m; \theta_s) + (1-c_{s})f_{\text{bg}}(m; \theta_{\text{bg}}).
\end{equation}
This distribution function incorporates the signal model, $f_s$, and the background model, $f_\text{bg}$, with the parameter $c_s$ controlling what fraction of the V$^{0}$ candidates are considered to be part of the signal. The parameters, $\theta$, include $c_s$ as well as the signal parameters, $\theta_{s}$, discussed in Section \ref{sec:sModels} and the background parameters, $\theta_\text{bg}$, which are the coefficients of the second degree polynomial. After obtaining $c_s$ from the fits, the raw yield of signal particles is calculated with:
$y^{\text{raw}} = c_s N_\text{V$^{0}$ Candidates}$.

\begin{figure*}[htbp]
\begin{center}
\includegraphics[width=0.8\textwidth]{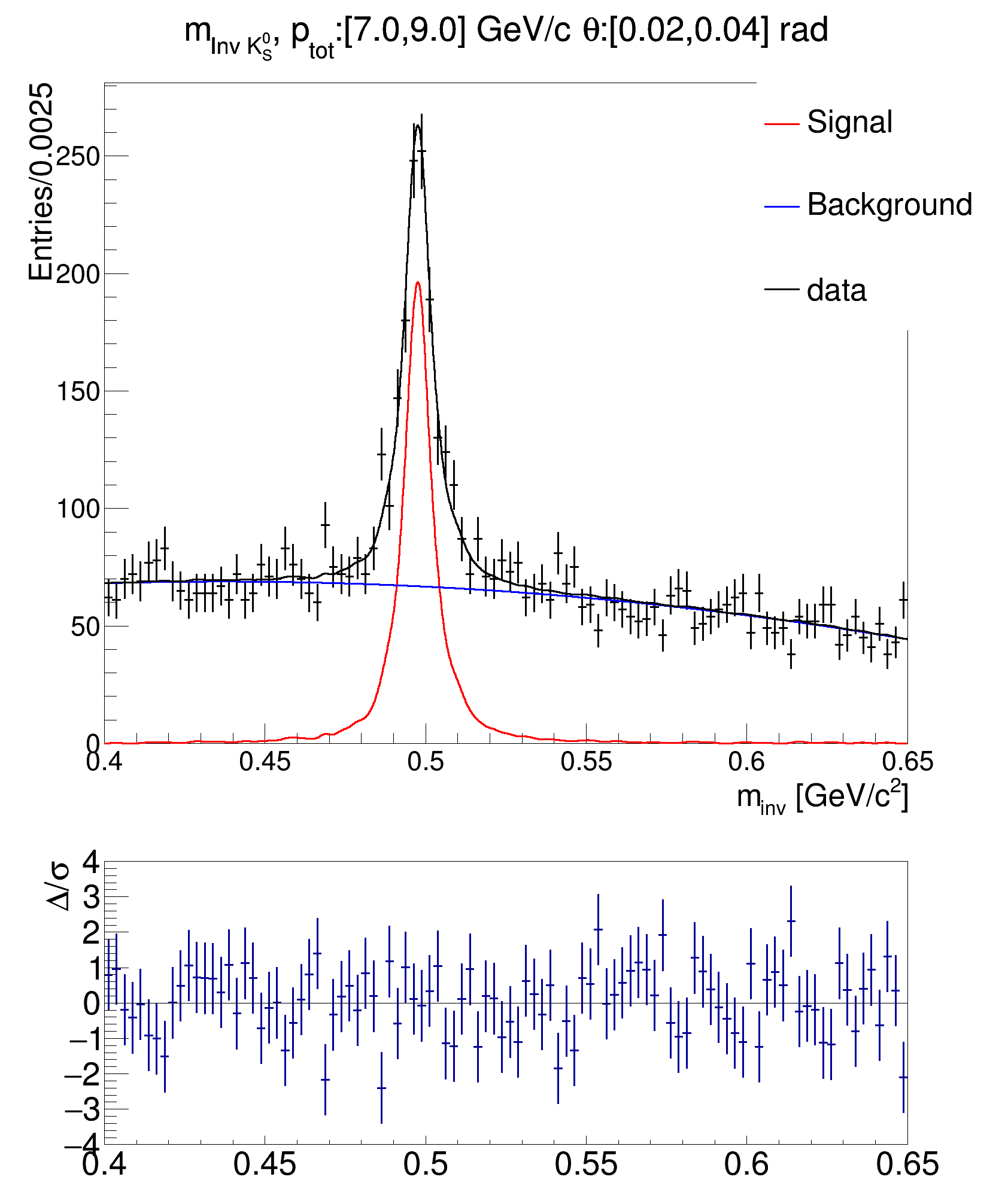}
\caption[Example \kos fit to data]{Example fit to the \kos invariant mass distribution in \piC data for an example kinematic bin. The $m_{inv}$ distribution and the fitted model is shown in the \textit{top}. The residuals of the fit are shown on the \textit{bottom}.
}
\label{fig:Fit_K0S_PiC60}
\end{center}
\end{figure*}

\begin{figure*}[htbp]
\begin{center}
\includegraphics[width=0.8\textwidth]{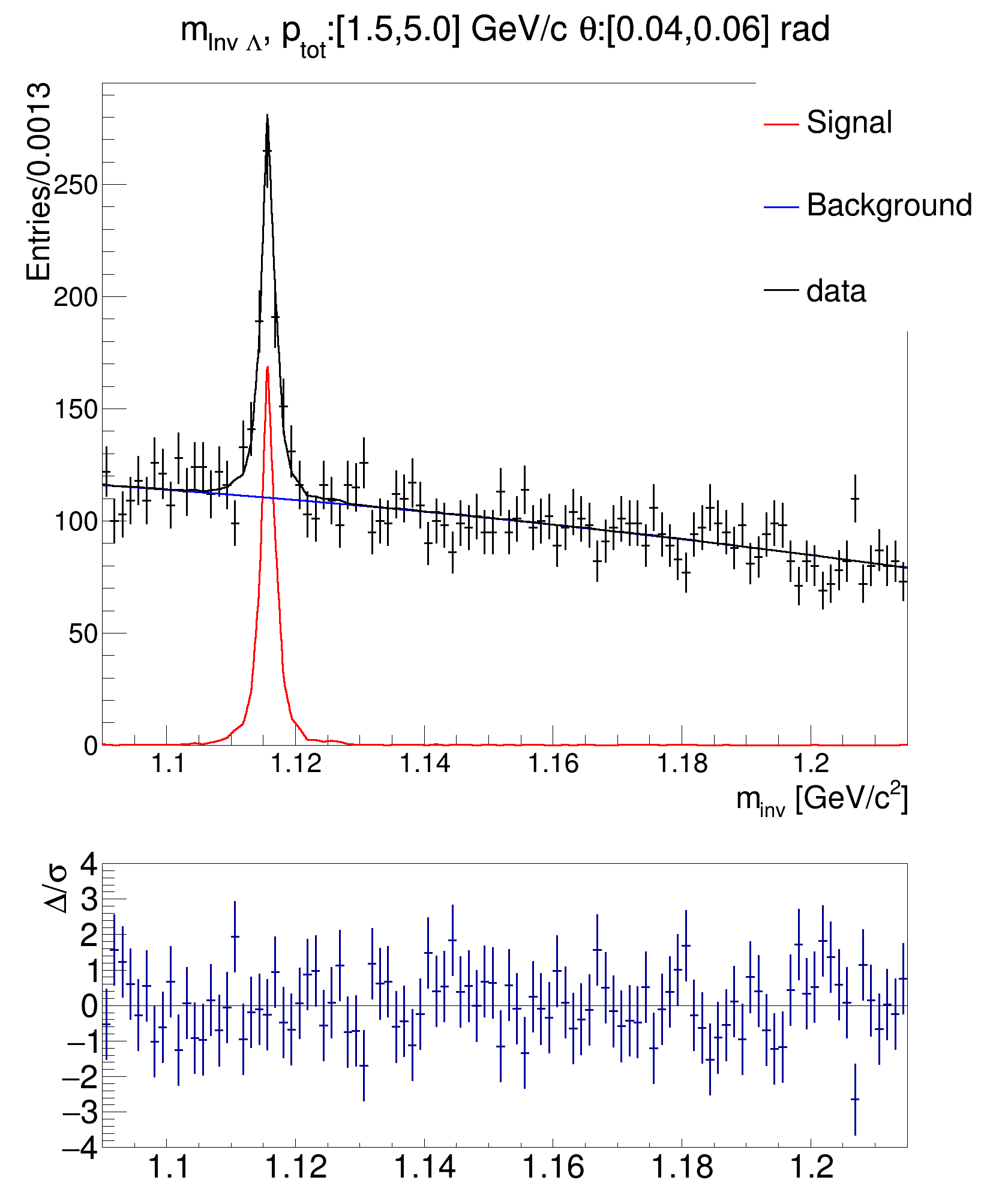}
\caption[Example \lam fit to data]{Example fit to the \lam invariant mass distribution in \piC data for an example kinematic bin. The $m_{inv}$ distribution and the fitted model is shown on the \textit{top}. The residuals of the fit are shown on the \textit{bottom}.
}
\label{fig:Fit_Lam_PiC60}
\end{center}
\end{figure*}

Figures~\ref{fig:Fit_K0S_PiC60} and \ref{fig:Fit_Lam_PiC60} show example fits to \kos and \lam invariant mass distributions from the \piC data set. Averaging over the fit results for all kinematic bins, the observed \kos mass was 498.7 \MeVcc, which is slightly higher than the known value of 497.6 \MeVcc~\cite{PhysRevD.98.030001}. The average of the widths of the invariant mass distributions was observed to be 17 \MeVcc. The \lam and \alam masses were both observed to be 1,117 \MeVcc, slightly higher than the known value of 1,116 \MeVcc~\cite{PhysRevD.98.030001}. The widths of the \lam and \alam distributions were found to be 6 \MeVcc and 7 \MeVcc, respectively. These small discrepancies in the masses compared to the known values are likely due to small biases in the momentum reconstruction of tracks.

\subsection{Corrections}

The raw yields obtained from the fits discussed in the previous section must be corrected for systematic effects. These can roughly be categorized into several effects: branching ratio of the decay, detector acceptance, feed-down corrections, reconstruction efficiency and selection efficiency. The combined effect of these individual effects can be estimated as a single correction factor from Monte Carlo simulations. Using \kos as an example, the correction factor for kinematic bin $i$ is given by:
\begin{equation}
c_i = \frac{N(\text{simulated \kos})}{N(\mbox{selected, reconstructed \kos})} = c_{\mbox{BR}} \times c_{\mbox{acc.}} \times c_{\mbox{feed-down}} \times c_{\mbox{rec. eff.}} \times c_{\mbox{sel. eff.}}. \label{eqn:MCCorrection}
\end{equation}
The correction factors are calculated in the analogous way for \lam and \alam. The correction factors are obtained from the MC production using the FTFP\_BERT physics list. 

\section{Analysis of Charged Hadron Spectra}
\label{sec:dEdx}

The analysis of produced charged hadrons is performed with a dE/dx analysis, which uses energy loss measured by the TPCs to separate particle species for both positively and negatively charged tracks. In particular, it was possible to measure spectra of produced \pip, \pim, \kp, \km and protons with this method. Compared to past analyses of interactions of 31~\GeVc protons with a thin carbon target~\cite{NA612007Pi,NA612007K,na61_t2k_thin}, in which the ToF-Forward wall was used, in this analysis, proton and kaon spectra were not able to be distinguished for certain momentum ranges on the basis of dE/dx information alone.

\subsection{Selection of Tracks}

The selection criteria are devised to remove off-time tracks and tracks coming from secondary interactions mistakenly reconstructed to the main interaction vertex. The selection cuts are also devised to filter out tracks with poorly determined track parameters, mainly $p$, $\theta$ and dE/dx. To start with, all tracks emanating from the main interaction vertex are considered for the dE/dx analysis. 

\subsubsection{Track Topologies}

There are a few ways tracks can be classified into different track topologies, including the initial direction of the tracks and which TPC chambers the tracks pass through. The most basic track topology classification used in NA61/SHINE analyses is the distinction between so-called right-side tracks (RSTs) and wrong-side tracks (WSTs) determined by the charge and direction emitted from the target. RSTs have a reconstructed $p_x$ that is in the same direction as the deflection by the vertex magnets. WSTs have a reconstructed $p_x$ opposite to the bending direction of the magnetic fields. This can be written more succinctly:
\begin{equation}
\begin{cases} 
p_x/q > 0 & \text{RST} \\
p_x/q < 0 & \text{WST} 
\end{cases}.
\end{equation}

For the same reconstructed momenta, RSTs and WSTs have very different detector acceptances, numbers of clusters and trajectories through different TPC sectors. Therefore, in this analysis, RSTs and WSTs undergo different selection criteria, are fit separately and had different corrections applied to them. This classification allows for a basic cross check, since these two samples lead to two somewhat independent measurements. For the purposes of this analysis, the distinction between RSTs and WSTs is not made for the first angular bin ([0,10] mrad for pions and [0,20] mrad for kaons and protons), because it is difficult to accurately distinguish between RSTs and WSTs near $\theta = 0$ mrad.

\subsubsection{Phi Cuts}

The azimuthal acceptance of the NA61/SHINE detector is highly dependent on the track topology and $\theta$. In order to obtain samples of tracks with similar numbers of clusters, $\phi$ cuts were devised as a function of $\theta$ bin and track topology and applied to the selection.

\subsubsection{Track Quality Cuts}

The impact parameter of tracks (distance from the main interaction vertex and the extrapolation of the track to the plane of the target) is required to be less than 2 cm in order to remove off-time tracks and tracks produced in secondary interactions. 

To ensure that the selected tracks have narrow enough dE/dx distributions to distinguish between particle species, at least 30 clusters are required in the VTPCs and MTPCs. In order to ensure tracks have good momentum estimations, there must be at least 4 clusters in the GTPC or 10 clusters in the VTPCs. Additionally, to remove tracks resulting from secondary interactions that were falsely reconstructed to the main interaction vertex, a cut is applied to tracks with no reconstructed GTPC and VTPC-1 clusters. This cut requires there to be fewer than 10 potential clusters in the VTPC-1 and fewer than 7 potential clusters in the GTPC, where the potential clusters are calculated by extrapolating tracks through the tracking system.

Several dE/dx cuts were applied to remove tracks with nonsensical dE/dx values (MIP) and rare heavier mass or doubly-charged particles:
\begin{equation}
\begin{cases}
0 < dE/dx < 2 & p \geq 2.2~\GeVc\\
0 < dE/dx < \langle dE/dx \rangle_{De} + 1 & p < 2.2~\GeVc
\end{cases}.
\end{equation}
These cuts remove much less than 1\% of tracks, so no correction is made to account for the dE/dx cuts. 

\begin{figure*}[htbp]
\begin{center}
\includegraphics[width=0.48\textwidth]{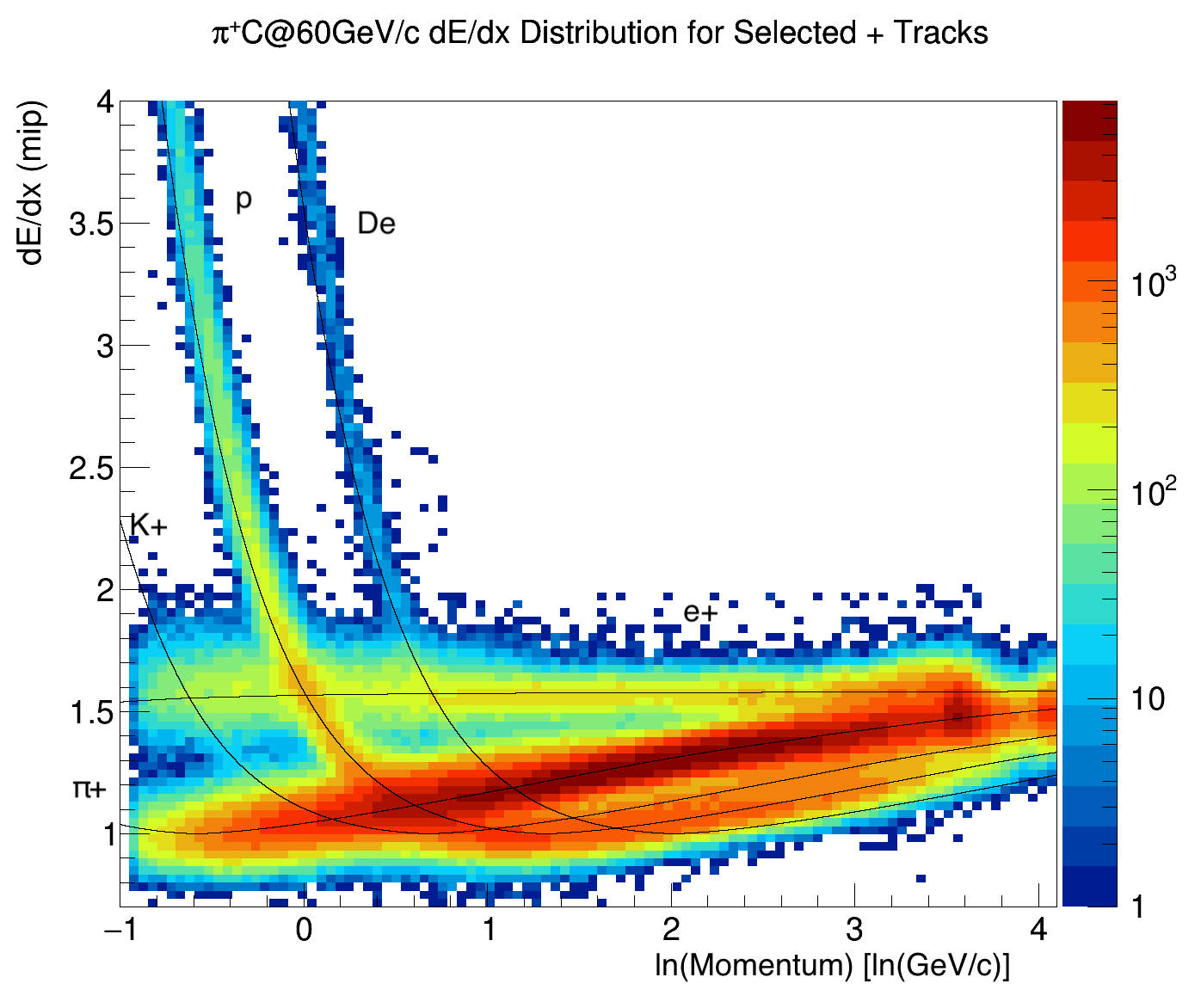}
\includegraphics[width=0.48\textwidth]{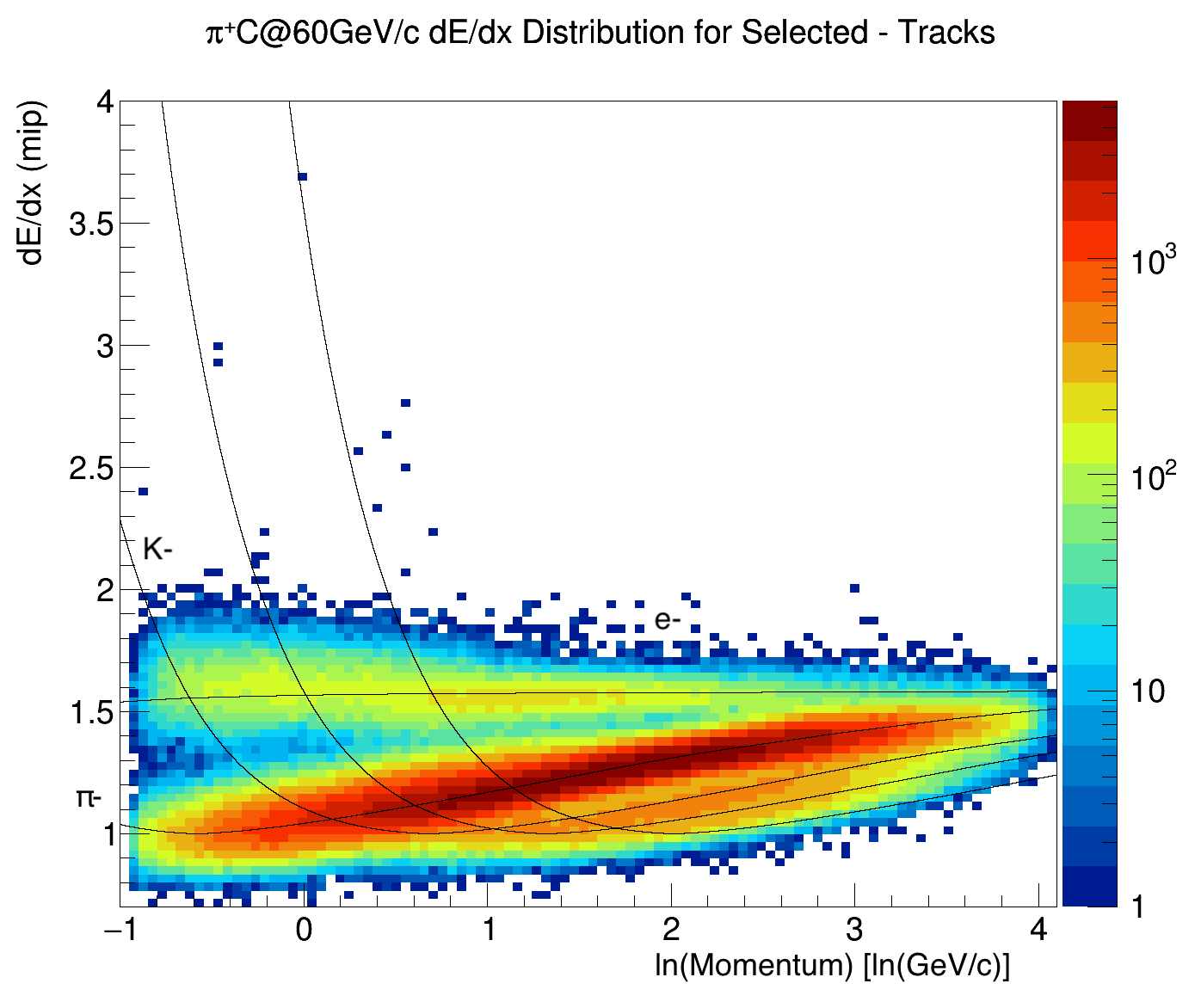}
\caption[dE/dx-Momenta distributions]{2-dimensional distributions of dE/dx and $p$ are shown for the selected positively (\textit{left}) and negatively (\textit{right}) charged tracks in the \piC analysis. The black lines represent the Bethe-Bloch predictions for the dE/dx mean position of electrons, pions, kaons, protons and deuterons.
}
\label{fig:dEdxP}
\end{center}
\end{figure*}

Figure~\ref{fig:dEdxP} shows the dE/dx-momentum distribution of the selected positively charged and negatively charged tracks. 

\subsection{Fitting to dE/dx Distributions}

For each analysis bin, a fit is used to determine the yields of each particle species. Five particles species and their anti-particles are considered: \epl, \pip, \kp, protons and deuterons. Positively charged and negatively charged tracks are simultaneously fit to better constrain the parameters. 

\subsubsection{dE/dx Model}

The mean dE/dx, $\langle\epsilon\rangle$, of charged particles passing through NA61/SHINE's TPCs depends on the particles' values of $\beta$, which, for particles of the same momentum, depend on their masses. A Bethe-Bloch table provides initial guesses of $\langle\epsilon\rangle$ for particle species within each bin.

The dE/dx distribution function describing the observed dE/dx of a charged particle passing through the TPCs depends on $\langle\epsilon\rangle$ and the distance traveled through the TPCs. The distribution closely resembles an asymmetric gaussian:
\begin{equation}
f(\epsilon,\sigma) = \frac{1}{\sqrt{2\pi\sigma}}\exp{\bigg[-\frac{1}{2}\Big(\frac{\epsilon-\mu}{\delta\sigma}\Big)^2\bigg]},
\end{equation}
where $\epsilon$ is the measured dE/dx of a track. The peak dE/dx of the distribution, $\mu$, is related to $\langle\epsilon\rangle$ through the relation:
\begin{equation}
\mu = \langle\epsilon\rangle -  \frac{4d\sigma}{\sqrt{2\pi}},
\end{equation}
where $d$ is the asymmetry parameter, which controls the asymmetry of the distribution through the relation:
\begin{equation}
\delta = \begin{cases} 1-d, & \mbox{if } \epsilon \leq \mu \\ 1+d, & \mbox{if } \epsilon > \mu \end{cases}.
\end{equation}

For a detector with uniform readout electronics, the width of the distribution for a single particle depends on the number of dE/dx clusters, $N_\text{Cl}$, and on $\langle\epsilon\rangle$: 
\begin{equation}
\sigma = \frac{\sigma_0 \langle \epsilon \rangle^{\alpha}}{\sqrt{N_{\text{Cl}}}},
\end{equation}
where the parameter, $\alpha$, controls how the width scales with $\langle\epsilon\rangle$ and $\sigma_0$ is the base dE/dx width of a single cluster. However, in NA61/SHINE, nonuniform readout electronics leads to different base widths for clusters reconstructed in different areas of the detector. This effect is most apparent in 3 main areas of the NA61/SHINE TPC system: the MTPCs, the two most upstream sectors of the VTPCs and the rest of the VTPCs. Different base widths characterizes each of these regions: $\sigma_\text{0, M}$, $\sigma_\text{0, Up}$ and $\sigma_\text{0, V}$. The dE/dx width of a single track can be parametrized more precisely by accounting for the numbers of clusters in each TPC region, $N_\text{Cl, Up}$, $N_\text{Cl, V}$ and $N_\text{Cl, M}$:
\begin{equation}
\sigma = \frac{\langle \epsilon \rangle^{\alpha}}{\sqrt{\frac{N_{\text{Cl, Up}}}{\sigma_{\text{0, Up}}^2} + \frac{N_{\text{Cl, V}}}{\sigma_{\text{0, V}}^2} + \frac{N_{\text{Cl, M}}}{\sigma_{\text{0, M}}^2}}}.
\end{equation}

At this point, some calibration and shape parameters need to be added in to account for imperfect dE/dx calibration, variation in pad response, variation in track angle and other effects that can cause $\langle\epsilon\rangle$ and $\sigma$ to deviate from the ideal model. Therefore, additional calibration parameters are added to allow the peaks and widths of the species distribution functions to vary slightly from the ideal model for each analysis bin. 

The full form of the single species distribution function is then:
\begin{equation}
f^{i,j}(\epsilon,p,N_\text{Cl, Up},N_\text{Cl, V},N_\text{Cl, M}) = \frac{1}{\sqrt{2\pi}\sigma^{i,j}_\text{cal}}\exp{\bigg[-\frac{1}{2}\Big(\frac{\epsilon-\mu^{i,j}_\text{cal}}{\delta\sigma^{i,j}_\text{cal}}\Big)^2\bigg]},
\end{equation}
where $\sigma^{i,j}_\text{cal}$ and $\mu^{i,j}_\text{cal}$ implicitly depend on the the momentum $p$, the number of clusters variables and the calibration parameters.

With these single-species distribution functions the single-track distribution functions can be built for both charges, $F^{+}$ and $F^{-}$:
\begin{equation}
F^{j}(\epsilon,p,N_\text{Cl, Up},N_\text{Cl, V},N_\text{Cl, M})= \sum_{i}{ y^{i,j} f^{i,j}(\epsilon,p,N_\text{Cl, Up},N_\text{Cl, V},N_\text{Cl, M})}
\end{equation}
where $y^{i,j}$ is the fractional contribution of species $i$ to the sample of tracks with charge $j$. The yields for each charge are constrained such that they sum to 1.

\subsubsection{Fitting Strategy}

\begin{figure*}[htbp]
\begin{center}
\includegraphics[width=0.9\textwidth]{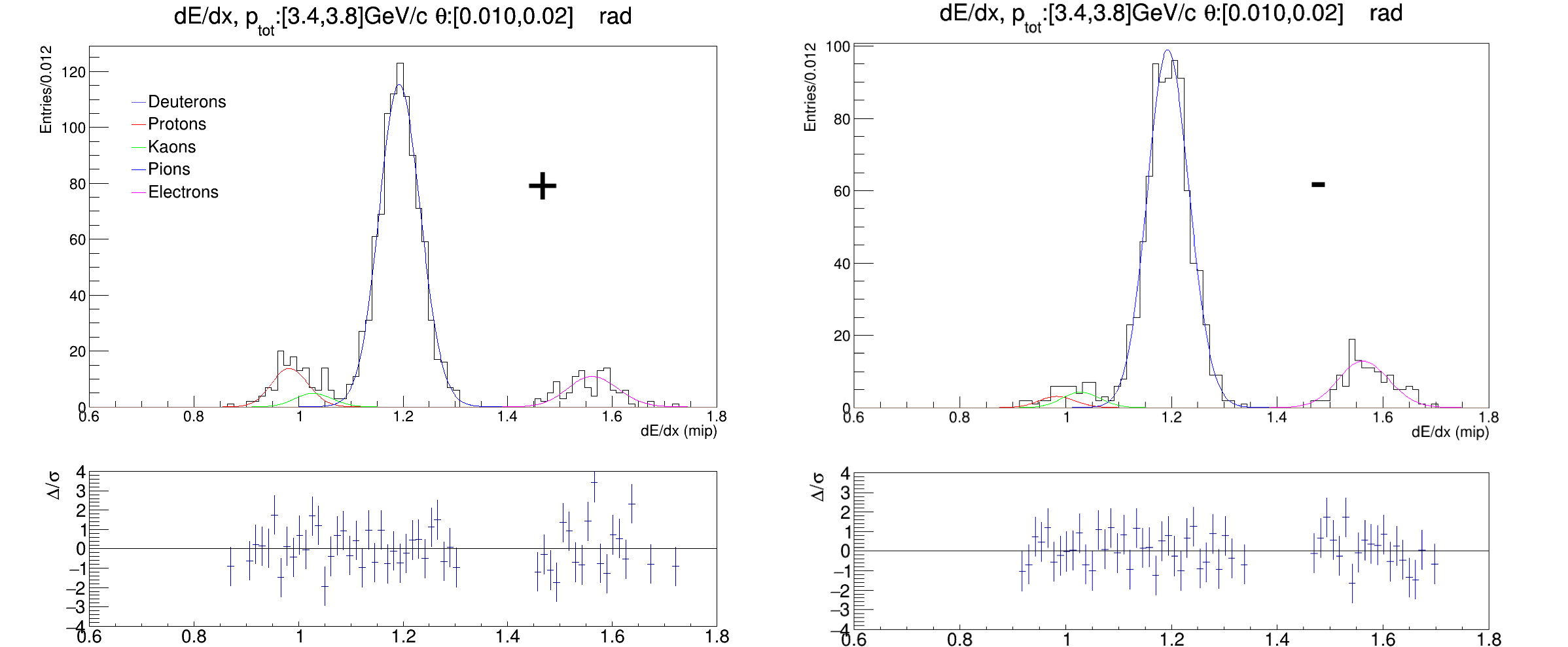}
\caption[Example fit to a dE/dx distribution from data]{An example fit to a dE/dx distribution is shown for the analysis of pions. On the \textit{top}, the dE/dx distributions are shown for positively charged tracks (\textit{left}) and negatively charged tracks (\textit{right}) along with the fitted contributions due to the 5 particle species considered. On the \textit{bottom}, the residuals of the fit with respect to the dE/dx distribution are shown.
}
\label{fig:exFit1}
\end{center}
\end{figure*}

To perform the minimization, a continuous log-likelihood function is constructed:
\begin{equation}
\log{L} = \sum_{\text{+tracks}}{\log{F^{+}(\epsilon,p,N_\text{Cl, Up},N_\text{Cl, V},N_\text{Cl, M} ; \theta)}} + \sum_{\text{-tracks}}{\log{F^{-}(\epsilon,p,N_\text{Cl, Up},N_\text{Cl, V},N_\text{Cl, M} ; \theta)}}.
\end{equation}
The log-likelihood function involves a sum over all of the positively and negatively charged tracks for a given analysis bin. In addition to the constraint that the yield fractions add up to 1 for each charge, soft constraints are applied to avoid the parameters converging to unreasonable values. For example, without constraints, it is easy for two species to swap the location of their dE/dx means. For fits to the target-removed data, all of the parameters are fixed to the fitted values from the target-inserted fits, except for the particle yields. Figure~\ref{fig:exFit1} shows a fit to the dE/dx distribution of an example bin. The estimated raw yield of a particle species in analysis bin $k$ is obtained by multiplying the fractional yield obtained from the fit, $y^{i,j}_k$, by the number of positively or negatively charged tracks in that bin, $N^{i}_k$:
\begin{equation}
Y^{i,j,\text{raw}}_k = y^{i,j }_k N^{i}_k.
\end{equation}
For each of the \pip, \pim, \kp, \km and proton analyses, a raw yield is obtained for each bin and for both the target-inserted and target-removed samples. 

\subsection{Corrections}

\subsubsection{Fit Bias Corrections}
\label{sec:fakeDEDX}

Simulated dE/dx distributions were generated in order to estimate the bias and the standard deviation of the particle yields obtained from the fitting procedure. 50 simulated dE/dx distributions for each analysis bin were built from the dE/dx model discussed in the previous section. The kinematic variables of tracks from data and the resulting hadron yields were taken as inputs for the dE/dx simulation. The fit parameters are varied according to the spread of fit results observed in data. 

The biases and standard deviations in the fitted yields are determined from the results of fits to these simulated dE/dx distributions. In general, the biases in the pion yields are small. The biases of the proton and kaon yields are larger in the high momentum regions and near the Bethe-Bloch crossing regions, where the particle distributions overlap significantly. The biases are used to correct the fit results with correction factors, $c^\text{fit}_k$, and the standard deviations are used to estimate the uncertainties related to the fitting procedure. 

\subsubsection{Monte Carlo Corrections}

The raw yields of particles obtained from the dE/dx fits must be corrected for a number of systematic effects. These can roughly be organized into: detector acceptance, feed-down corrections, reconstruction efficiency, selection efficiency and in the case of pions, muon contamination. The combined effect of these individual effects can be estimated as an overall correction factor from Monte Carlo simulations, as was done in the V$^{0}$ analysis. A few of the forward kinematic bins contain particle trajectories that strike the S4. A further correction was applied to account for this effect, which reached about 7\% for a few of the \pip bins, but did not exceed 2\% for the other charged hadron species.

In the case of corrections for \pip and \pim, because the dE/dx signal from muons is indistinguishable from pions, muon tracks that pass the selection criteria and are fitted to the main interaction vertex must also be accounted for:
\begin{equation}
c^\text{MC}_k = \frac{N(\text{produced, simulated \pipm})_k}{N(\text{selected, reconstructed \pipm, \mupm})_k} = c_{\text{acc.}} \times c_{\text{feed-down}} \times c_{\text{rec. eff.}} \times c_\text{sel. eff.} \times c_{\text{\mupm}}.
\end{equation}

\subsubsection{Feed-down Reweighting}

The feed-down correction, which can be as large as 20\% for protons, is the main component of the MC correction factor that depends on the physics model. We cannot assume that the production of \lam, \alam and \kos is accurately predicted by the physics generators. This incurs an uncertainty on the MC corrections and subsequently, on the resulting multiplicity measurements.

We can constrain this uncertainty by reweighting our MC productions with the results of the V$^{0}$ analyses. When counting the number of reconstructed pions and protons passing the selection criteria, a weight is applied whenever that reconstructed track comes from a \kos, \lam or \alam:
\begin{equation}
w_\beta = \frac{m^{\text{data}}_\beta}{m^{\text{MC}}_\beta},
\end{equation}
where $m^{\text{data}}_\beta$ is the multiplicity measured in bin $\beta$ of the V$^{0}$ analysis and $m^{\text{MC}}_\beta$ is the multiplicity observed in the simulation in that bin. 

\section{Systematic Uncertainties on Spectra Measurements}
\label{sec:spectrauncert}

A number of possible systematic effects on the multiplicity measurements have also been evaluated. These include biases and uncertainties incurred by the fitting procedures, uncertainties associated with the MC corrections, uncertainties incurred in the selection procedures and uncertainties associated with the reconstruction. On top of the uncertainties described in the following sections, an overall normalization uncertainty is attributed to all of the multiplicity measurements. It has been estimated to be $\pm^{2}_{1}$\% by propagating the uncertainties on the normalization constants derived from the integrated cross section analysis through the multiplicity calculation, which will be discussed in Section \ref{sec:results}.

\subsection{Fit Model Uncertainty}

In the V$^{0}$ analysis, it cannot be assumed that the fits to the invariant mass distributions perfectly separate the signal from the background. To check for biases in the fit results, the fitting procedure is performed on additional MC productions using GEANT4 physics lists QGSP\_BERT, QBBC and FTF\_BIC. With these samples, the numbers of true \kos, \lam and \alam are known, so the bias and the standard deviation of the fit result can be calculated. For \kos, \lam and \alam, the fitting bias, $\mu$, on the signal fraction, $c_s$, was found to be $3.3\% \pm 2.7\%$, $4.8\% \pm 4.2\%$ and $11\% \pm 10\%$, respectively. The bias is not used as a correction for the fit results, but the values of $\mu \pm \sigma$ are taken as upper and lower uncertainties on the signal fraction, which are propagated through the multiplicity calculation. 

The fit model uncertainties on the charged spectra are obtained from the fits to simulated dE/dx distributions discussed in Section~\ref{sec:fakeDEDX}. The standard deviations in the particle yields are propagated to the multiplicities and taken as the uncertainties associated with the fitting routine.

\subsection{Physics Uncertainties}

Assuming different underlying physics can lead to different MC correction factors. For example, if the acceptance changes as a function of $p$ and $\theta$, different MC-predicted $p$ and $\theta$ distributions can lead to different MC correction factors. This uncertainty is evaluated by applying correction factors obtained with additional MC productions using the physics lists: QGSP\_BERT, QBBC and FTF\_BIC. The upper and lower bounds on the uncertainties are taken as the maximum and minimum values of the multiplicity obtained using these additional MC correction factors for each analysis bin.

\subsection{Feed-down Uncertainties}

The MC corrections account for a background of produced hadrons coming from heavier weakly-decaying particles. However, it cannot be assumed that the physics generators correctly predict the production rates of these heavier weakly-decaying hadrons. This uncertainty is evaluated by assuming a 50\% uncertainty on the number of reconstructed feed-down particles when calculating the MC correction factors, unless the feed-down particle was a reweighted \kos, \lam or \alam. In this case, the upper and lower uncertainties on the associated neutral hadron spectra are assigned to the weight assigned to the feed-down particles. These uncertainties are then propagated to the multiplicities. This reweighting treatment results in a significant reduction of the uncertainties on the \pip, \pim and proton spectra. 

\subsection{Selection Uncertainties}

Although the MC corrections account for the efficiency of the selection cuts, differences in data and MC could incur systematic biases in the result. It was found that tracks in data are typically composed of around 5\% fewer clusters than tracks in MC for the same kinematics. To estimate the selection uncertainty, alternative sets of MC corrections were obtained by artificially decreasing the numbers of clusters in MC tracks by 5\%. Higher multiplicities are obtained when applying these alternative correction factors, which are taken as the upper bounds of the selection uncertainty.

\subsection{Reconstruction Uncertainties}

The MC corrections should account for inefficiencies in the reconstruction of tracks and V$^{0}$s if the geometry and detector response are perfectly modeled by the simulation. Differences between the real detector and the simulated detector could lead to systematic effects on reconstruction efficiency component of the MC corrections. To estimate this uncertainty, the detectors were purposefully moved in the detector description model used by the reconstruction. Specifically, eight alternative productions were made after shifting the VTPC-1 and VTPC-2 by +.2 mm and -.2 mm in the x direction and +.5 mm and -.5 mm in the y direction. These shifts are considered to be rather large when compared to the alignment effects seen in the calibration of the data.

The numbers of selected charged tracks and V$^{0}$ candidates were calculated from these alternative productions. The maximum difference in the number of candidate tracks/V$^{0}$s among the productions are calculated for the x shifts and the y shifts in each analysis bin. The effects of the x and y shifts are then added in quadrature to estimate the uncertainty for each bin. The resulting uncertainties are generally less than 1\% and do not exceed 4\%.

\subsection{Momentum Uncertainties}

There is an uncertainty on the reconstruction of momentum due to uncertainties in converting the magnet currents to magnetic field strength. This uncertainty can be investigated by checking the invariant mass distributions fitted in the V0 analysis. The variation in the fitted means of the invariant mass distributions of \kos and \lam indicate an uncertainty in the reconstruction of momentum of up to 0.3\%. Uncertainties on the measured multiplicities due to misreconstructed momenta was determined by varying the momenta of tracks by 0.3\% and recalculating the numbers of selected tracks and V$^{0}$ candidates. This uncertainty was determined to be less than 1\% for the majority of the analysis bins, but is on the level of the statistical uncertainty for some of the analysis bins at the edges of the phase space measured.

\subsection{Breakdowns in Uncertainties}

\begin{figure*}[p]
\begin{center}
\includegraphics[width=0.89\textwidth]{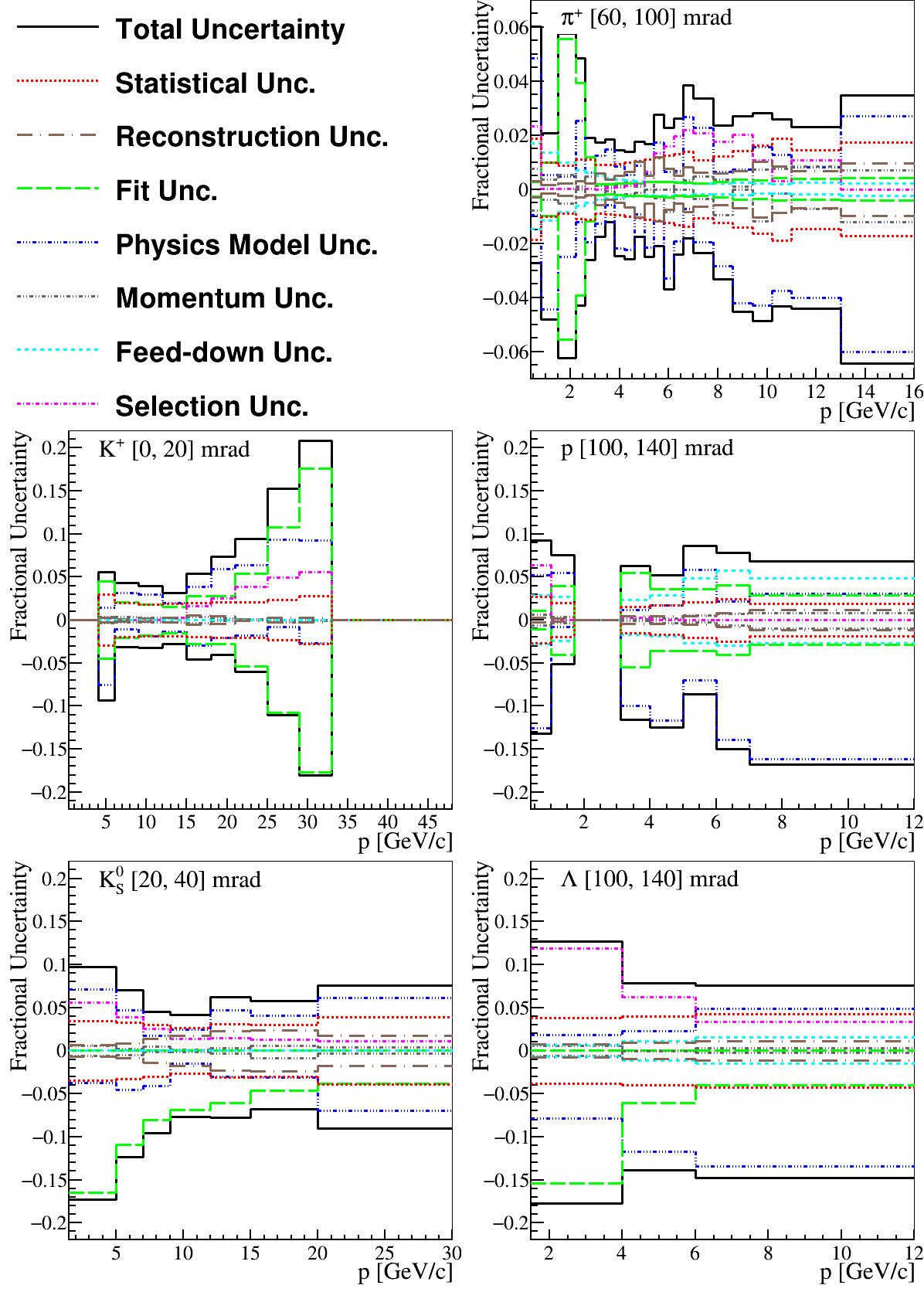}
\caption[1D \kos uncertainties breakdown]{The breakdown of the fractional uncertainties on \pip, \kp, proton, \kos and \lam spectra from \piC interactions for select representative angular bins. The upper and lower uncertainties are shown on the positive and negative sides of the y axes.
}
\label{fig:selectedUnc}
\end{center}
\end{figure*}

The breakdowns in the uncertainties for \pip, \kp, proton, \kos and \lam spectra from \piC interactions are shown for representative angular bins in Figure~\ref{fig:selectedUnc}. These breakdowns include statistical uncertainties, fit uncertainties, physics uncertainties, feed-down uncertainties, selection uncertainties momentum uncertainties and reconstruction uncertainties. The breakdowns of the uncertainties are largely similar for the measured hadron spectra from interactions of \piBe. Figures in Ref. \cite{edmsTables} present breakdowns of the uncertainties for the complete set of spectra measurements for interactions of \piC and \piBe. 

For the neutral spectra, the uncertainties are within 10\% in the kinematic regions with good detector acceptance and high statistical power. In the low-momentum regions, uncertainties associated with the fitting routine tend to dominate the lower uncertainties and selection uncertainties tend to dominate the upper uncertainties. The physics model uncertainty is typically the largest component of the uncertainty in the high momenta regions. 

For the charged spectra, the total uncertainties are generally around 5\% or less except in the kinematic regions with poor acceptance or poor dE/dx separation. In spectra of \pip, the largest uncertainties tend to be reconstruction uncertainties at high momenta and dE/dx fit uncertainties at low momenta. In the case of \pim, dE/dx fit uncertainties, physics model uncertainties and statistical uncertainties contribute the most to the total uncertainty. For kaons, dE/dx fit uncertainties are dominant in the majority of the phase space measured. For protons, uncertainties related to the physics model and dE/dx fit uncertainties are dominant for the majority of the phase space measured.

\section{Differential Production Multiplicity Measurements}
\label{sec:results}

The differential production multiplicity is the yield of particles produced per production interaction per unit momentum per radian in each kinematic bin $k$.
The production multiplicity for neutral hadrons can be written:
\begin{equation}
\frac{d^2n_k}{dpd\theta} = \frac{\sigma_{\text{trig}}c^{MC}_{k}}{f_{\text{prod}}\sigma_{\text{prod}}(1-\epsilon)\Delta p \Delta\theta}\bigg(\frac{Y_{k}^{I}}{N^I} - \frac{\epsilon Y_{k}^{R}}{N^R}\bigg)\label{eqn:multiplicity},
\end{equation}
where $\Delta p \Delta\theta$ is the size of bin $k$, and the yields, $Y^{I,R}_k$, are the total numbers of particles observed in bin $k$ determined by the invariant mass fits for target-inserted and target-removed data. The constants $\sigma_{\text{trig}}$, $\sigma_{\text{prod}}$, $f_\text{prod}$ and $\epsilon$ are determined from the integrated cross section analysis and $N^I$ and $N^R$ are the numbers of selected events with the target inserted and target removed. The differential cross section is related to the multiplicity by a factor of $\sigma_{\text{prod}}$:
\begin{equation}
\frac{d^2\sigma_k}{dpd\theta} = \sigma_{\text{prod}}\frac{d^2n_k}{dpd\theta}.
\end{equation}

In order to calculate the multiplicity for produced charged hadrons (for each track topology - RST and WST), an additional correction factor is required for the fit bias corrections, $c^\text{fit}$:
\begin{equation}
m_k = \frac{d^2n_k}{dpd\theta} = \frac{\sigma_{\text{trig}}c^\text{MC}_k c^\text{fit}_k}{f_{\text{prod}}\sigma_{\text{prod}}(1-\epsilon)\Delta p \Delta\theta}\bigg(\frac{Y_{k}^{I}}{N^I} - \frac{\epsilon Y_{k}^{R}}{N^R}\bigg).
\end{equation}

For kinematic bins for which the detector acceptance and fit reliability is sufficient enough for multiplicity measurements in both RST and WST bins, the single-side multiplicities, $m_\text{R}$ and $m_\text{W}$, are merged by taking the weighted average:
\begin{equation}
m_\text{merged} = \sigma^{2}_\text{merged}\bigg(\frac{m_\text{R}}{\sigma^{2}_\text{R}}+\frac{m_\text{W}}{\sigma^{2}_\text{W}}\bigg),
\end{equation}
where the merged uncertainty, $\sigma_{merged}$ is calculated with:
\begin{equation}
\frac{1}{\sigma^{2}_\text{merged}} = \frac{1}{\sigma^{2}_\text{R}} + \frac{1}{\sigma^{2}_\text{W}}.
\end{equation}
The uncertainties on the individual RST and WST multiplicities consider both the statistical uncertainties and the fit uncertainties:
\begin{equation}
\sigma_\text{R,W} = \sqrt{\sigma^{2}_\text{R,W stat} + \sigma^{2}_\text{R,W fit}}.
\end{equation}
In analysis bins for which the detector acceptance is only sufficient for either RSTs or WSTs, only the single-side multiplicity and uncertainty is taken as the result.

\begin{figure*}[p]
\begin{center}
\includegraphics[width=0.9\textwidth]{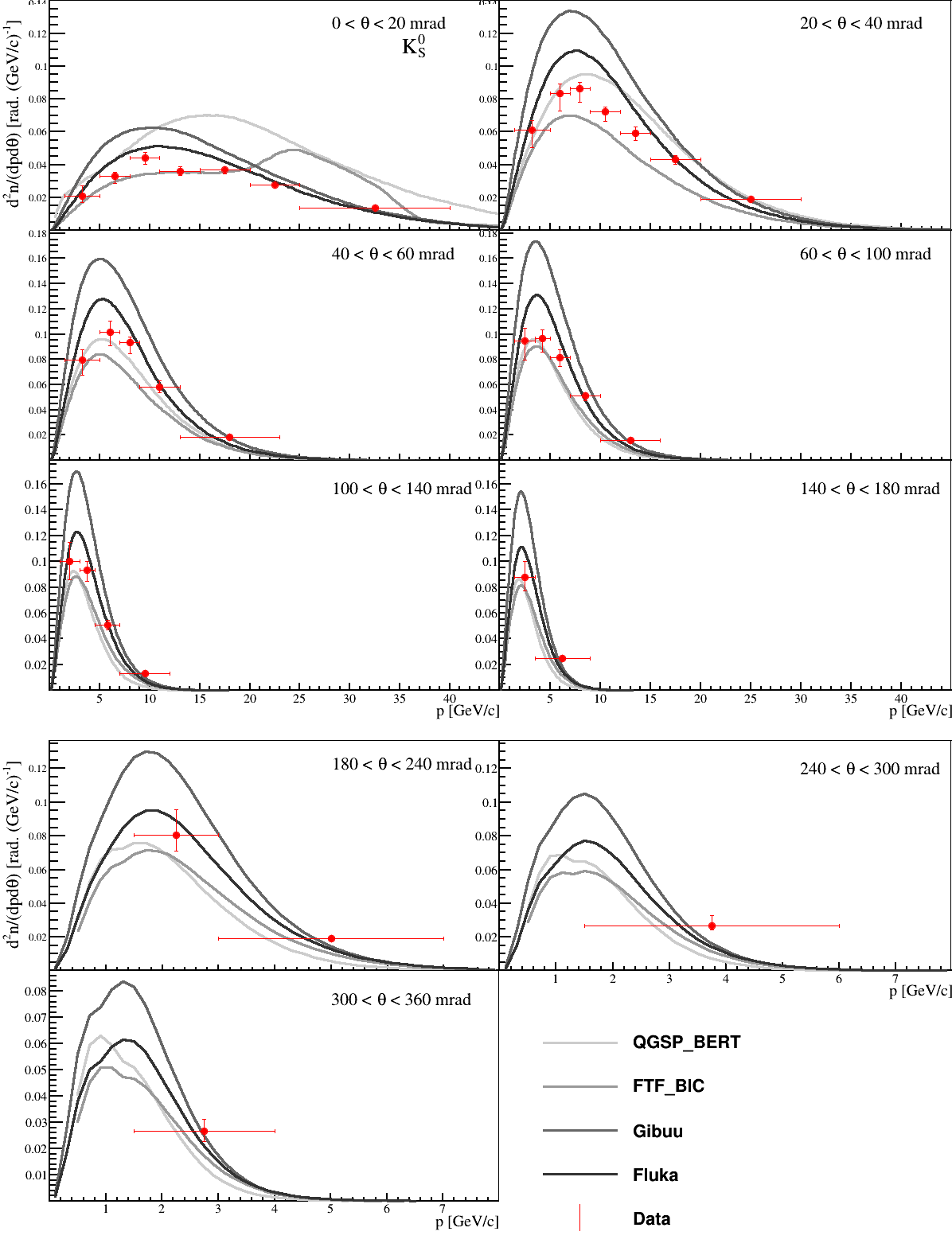}
\caption[1D \kos spectra]{\kos multiplicity spectra from \piC interactions are shown for different regions of $\theta$. The error bars represent total uncertainties except for the normalization uncertainty. The results are compared to the predictions of the GEANT4 physics lists QGSP\_BERT and FTF\_BIC as well as GiBUU2019 and FLUKA2011.
}
\label{fig:K0S_PiC60_1D}
\end{center}
\end{figure*}

\begin{figure*}[p]
\begin{center}
\includegraphics[width=0.9\textwidth]{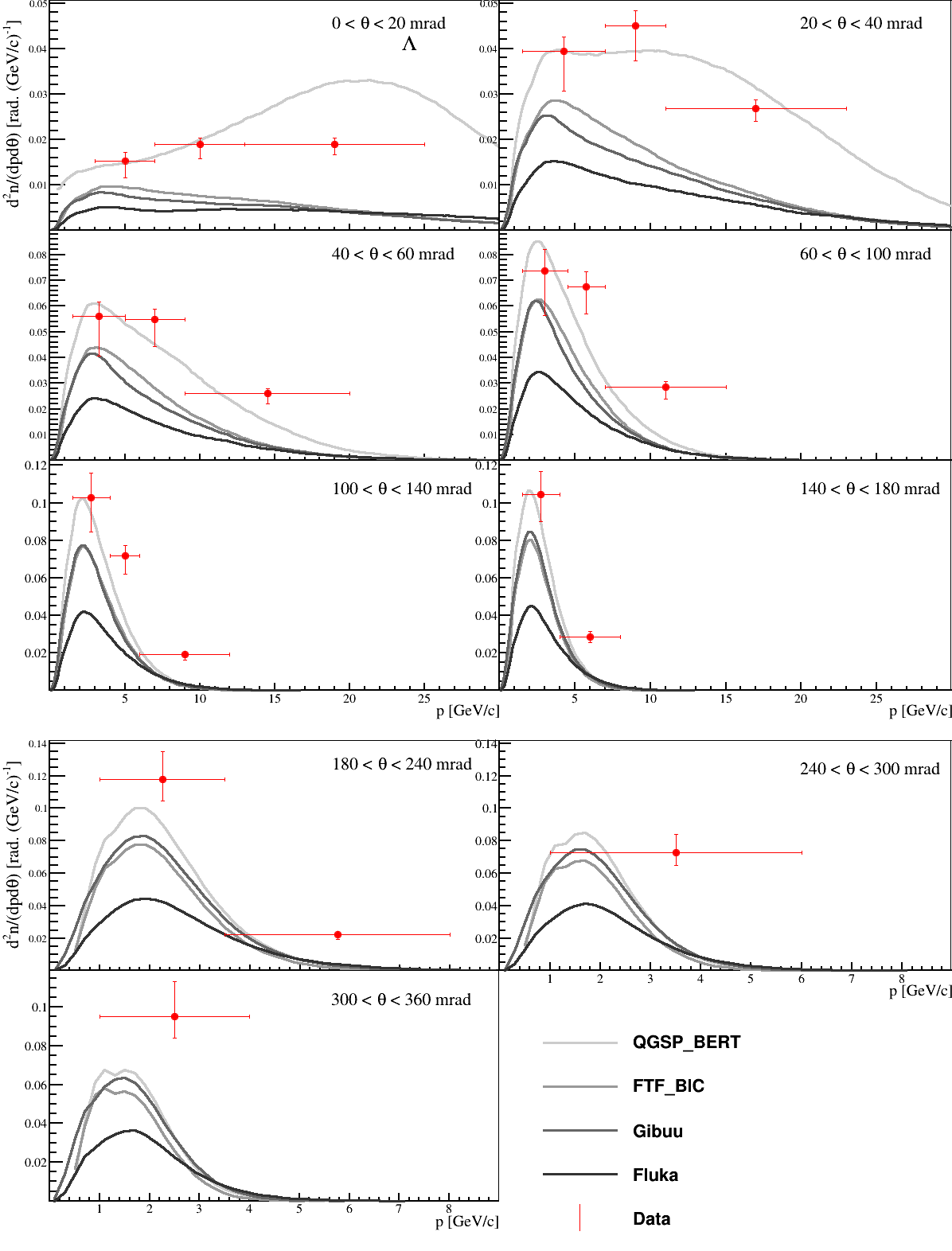}
\caption[1D \lam spectra]{\lam multiplicity spectra from \piC interactions are shown for different regions of $\theta$. The error bars represent total uncertainties except for the normalization uncertainty. The results are compared to the predictions of the GEANT4 physics lists QGSP\_BERT and FTF\_BIC as well as GiBUU2019 and FLUKA2011.
}
\label{fig:Lam_PiC60_1D}
\end{center}
\end{figure*}

\begin{figure*}[p]
\begin{center}
\includegraphics[width=0.9\textwidth]{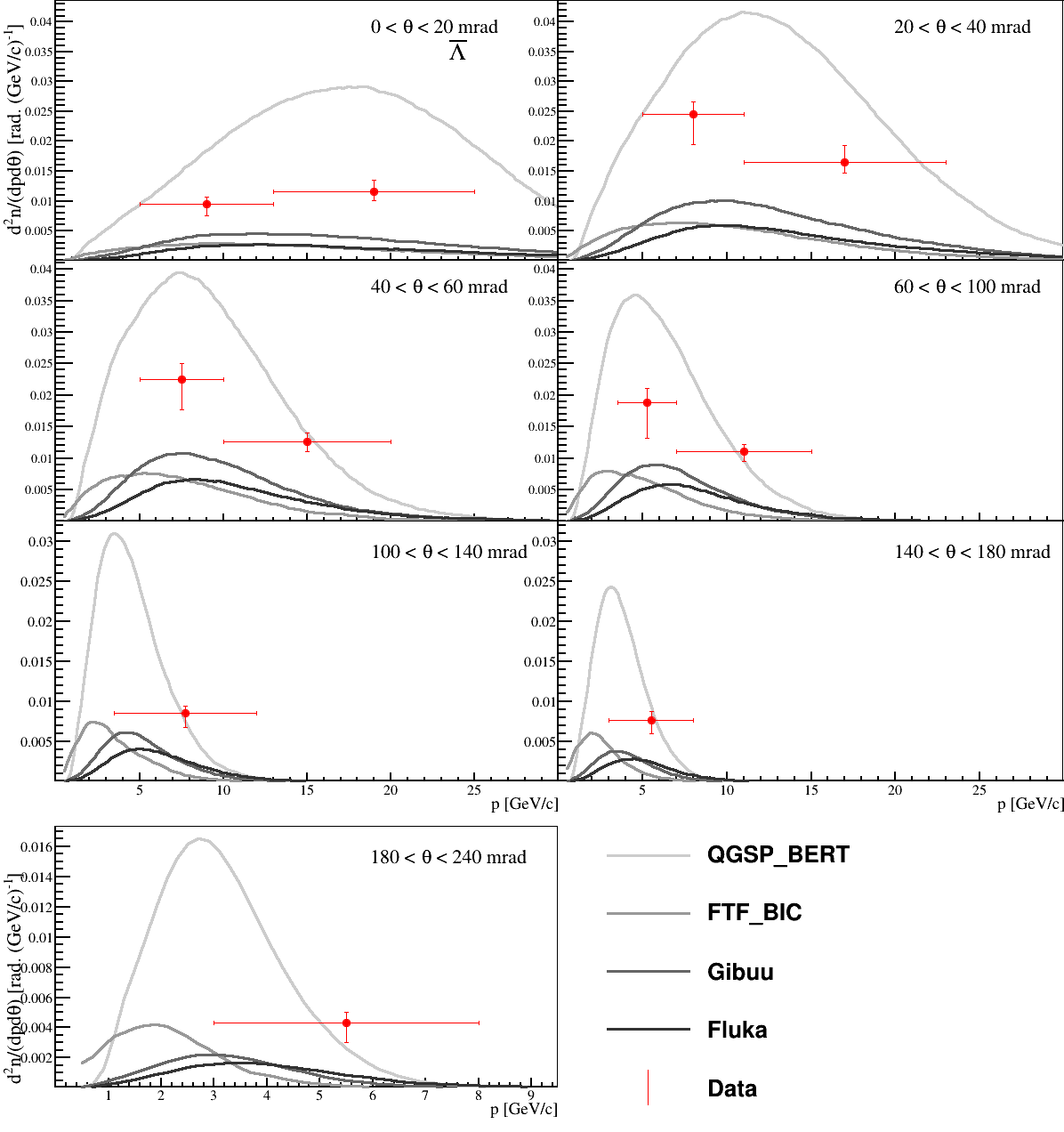}
\caption[1D \alam spectra]{\alam multiplicity spectra from \piC interactions are shown for different regions of $\theta$. The error bars represent total uncertainties except for the normalization uncertainty. The results are compared to the predictions of the GEANT4 physics lists: QGSP\_BERT and FTF\_BIC as well as GiBUU2019 and FLUKA2011.
}
\label{fig:ALam_PiC60_1D}
\end{center}
\end{figure*}

Multiplicity spectra obtained for \kos, \lam and \alam in \piC interactions are presented in Figures~\ref{fig:K0S_PiC60_1D}, \ref{fig:Lam_PiC60_1D} and \ref{fig:ALam_PiC60_1D}. The spectra are shown as 1-dimensional momentum spectra for individual bins of $\theta$. The error bars represent the total uncertainty except for the normalization uncertainty. The results are compared to the predictions of the GEANT4 physics lists: QGSP\_BERT and FTF\_BIC as well as GiBUU2019~\cite{Buss:2011mx} and FLUKA2011.2x.7~\cite{Battistoni:2015epi,Bohlen:2014buj,Ferrari:2005zk}. In general, the \kos spectra fall within the range of predictions of the models used. No single model describes the \kos spectra aptly for the full phase space. The models exhibit a large variability in their predictions of \lam and especially \alam spectra. QGSP\_BERT seems to provide the best prediction of \lam spectra, while no single model seems to provide a satisfactory description of \alam spectra. Tables in Ref. \cite{edmsTables} present the numerical values of the multiplicity measurements of \kos, \lam and \alam along with statistical, systematic and total uncertainties for each kinematic bin analyzed. The normalization uncertainty of $\pm^{2}_{1}$\% is not included in the values of the uncertainties shown in these tables but should be attributed to the multiplicity spectra of all hadron species analyzed.

\begin{figure*}[p]
\begin{center}
\includegraphics[width=0.85\textwidth]{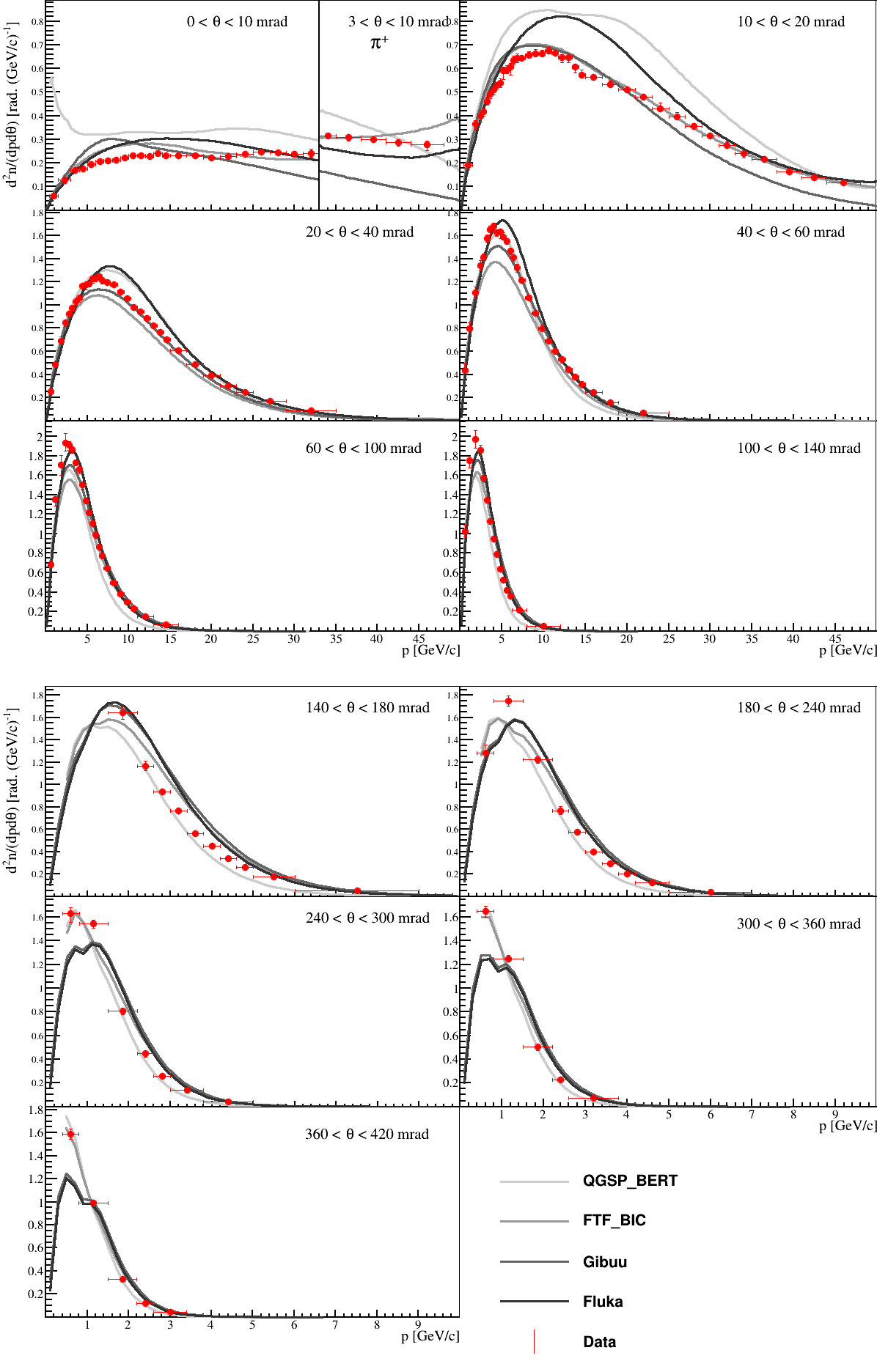}
\caption[\pip spectra]{\pip multiplicity spectra from \piC interactions are shown. The error bars represent total uncertainties except for the normalization uncertainty. Note that the first angular bin ([0,10] mrad) is divided into two regions. For momenta less than 33~\GeVc, the angular range is [0,10] mrad and for momenta greater than 33~\GeVc, the angular range is [3,10] mrad. The results are compared to the predictions of the GEANT4 physics lists: QGSP\_BERT and FTF\_BIC as well as GiBUU2019 and FLUKA2011.
}
\label{fig:PiP_PiC60_1D}
\end{center}
\end{figure*}

\begin{figure*}[p]
\begin{center}
\includegraphics[width=0.85\textwidth]{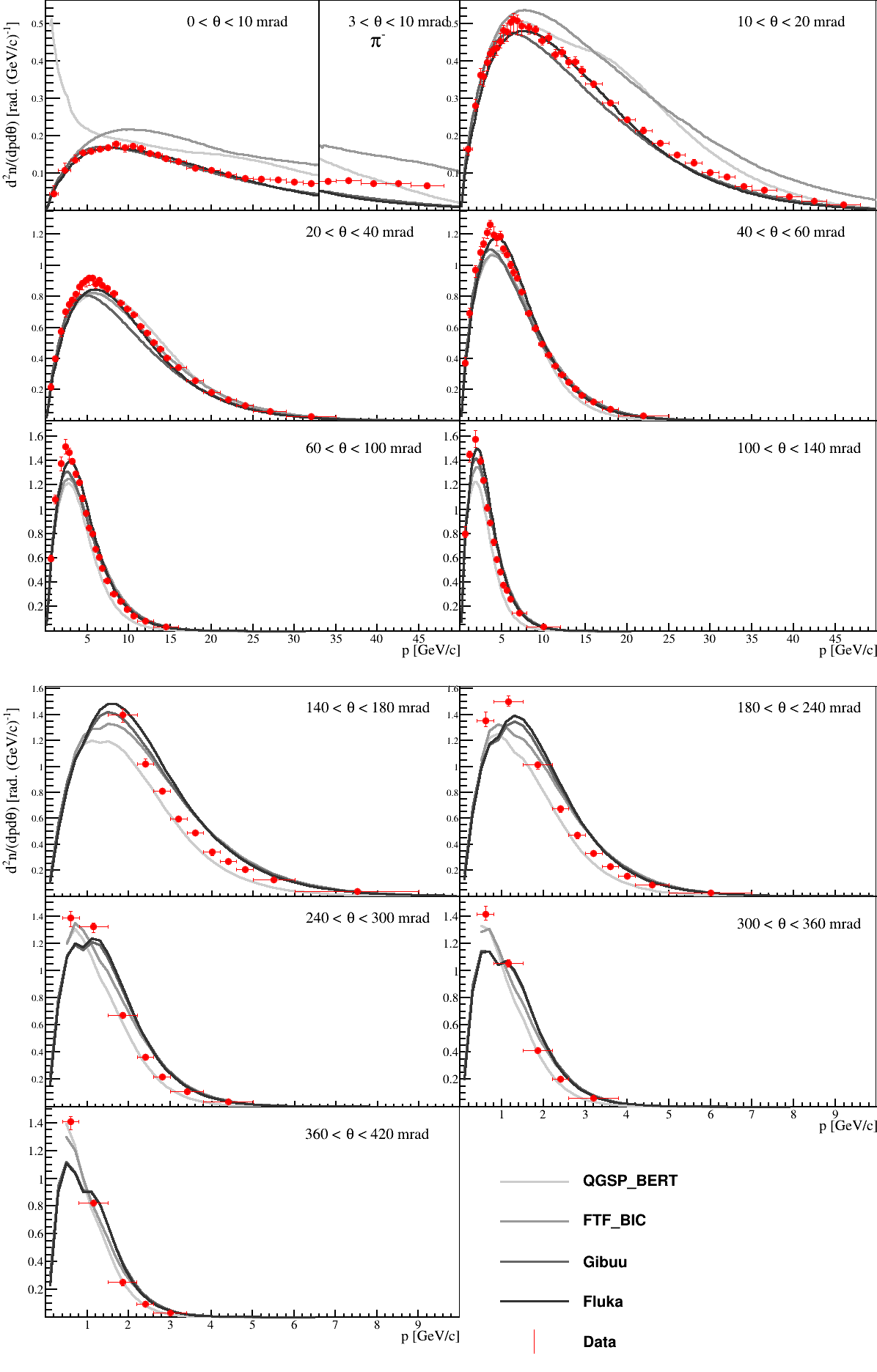}
\caption[\pim spectra]{\pim multiplicity spectra from \piC interactions are shown. The error bars represent total uncertainties except for the normalization uncertainty. Note that the first angular bin ([0,10] mrad) is divided into two regions. For momenta less than 33~\GeVc, the angular range is [0,10] mrad and for momenta greater than 33~\GeVc, the angular range is [3,10] mrad. The results are compared to the predictions of the GEANT4 physics lists: QGSP\_BERT and FTF\_BIC as well as GiBUU2019 and FLUKA2011.
}
\label{fig:PiM_PiC60_1D}
\end{center}
\end{figure*}

\begin{figure*}[p]
\begin{center}
\includegraphics[width=0.85\textwidth]{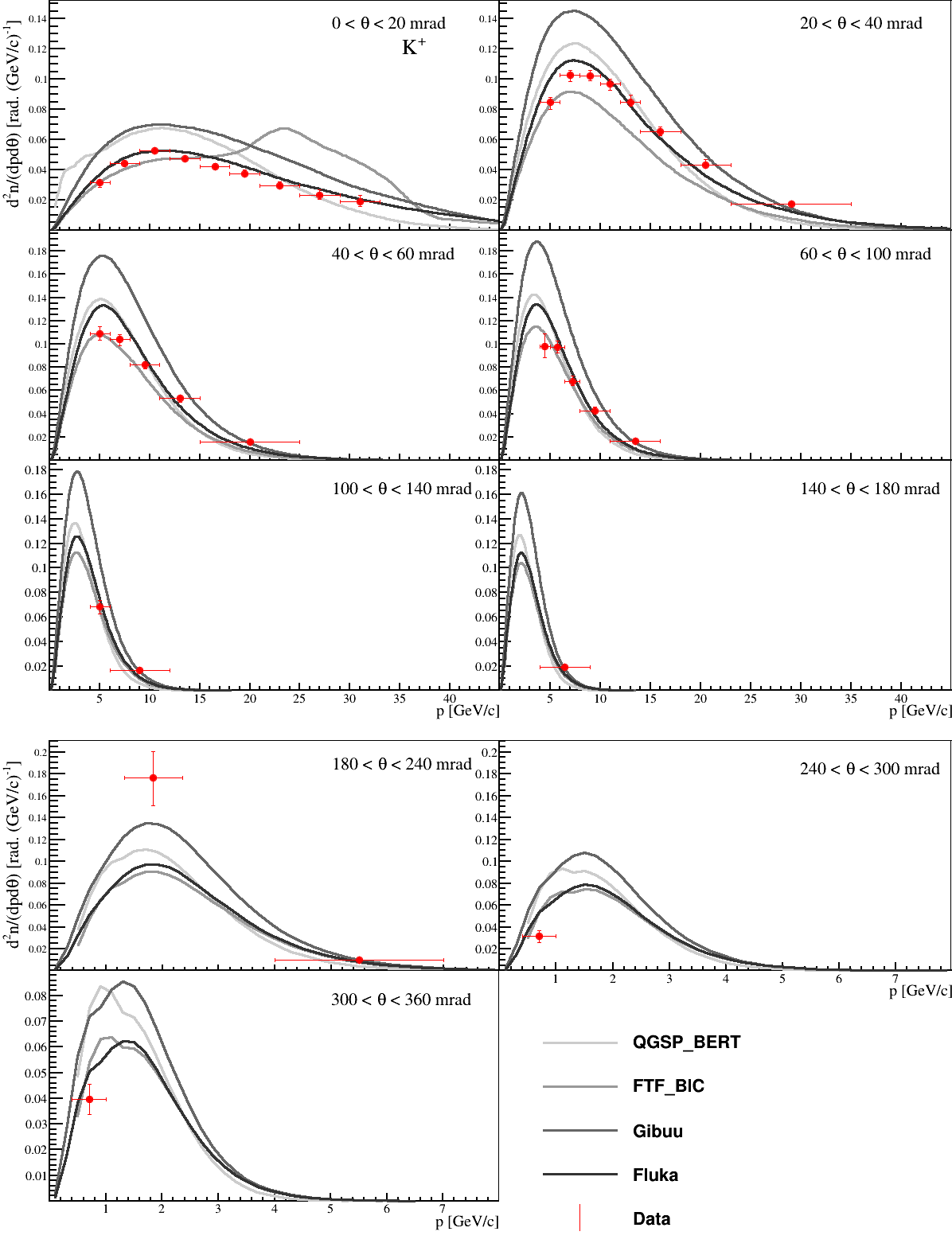}
\caption[\kp spectra]{\kp multiplicity spectra from \piC interactions are shown. The error bars represent total uncertainties except for the normalization uncertainty. The results are compared to the predictions of the GEANT4 physics lists: QGSP\_BERT and FTF\_BIC as well as GiBUU2019 and FLUKA2011.
}
\label{fig:KP_PiC60_1D}
\end{center}
\end{figure*}

\begin{figure*}[p]
\begin{center}
\includegraphics[width=0.85\textwidth]{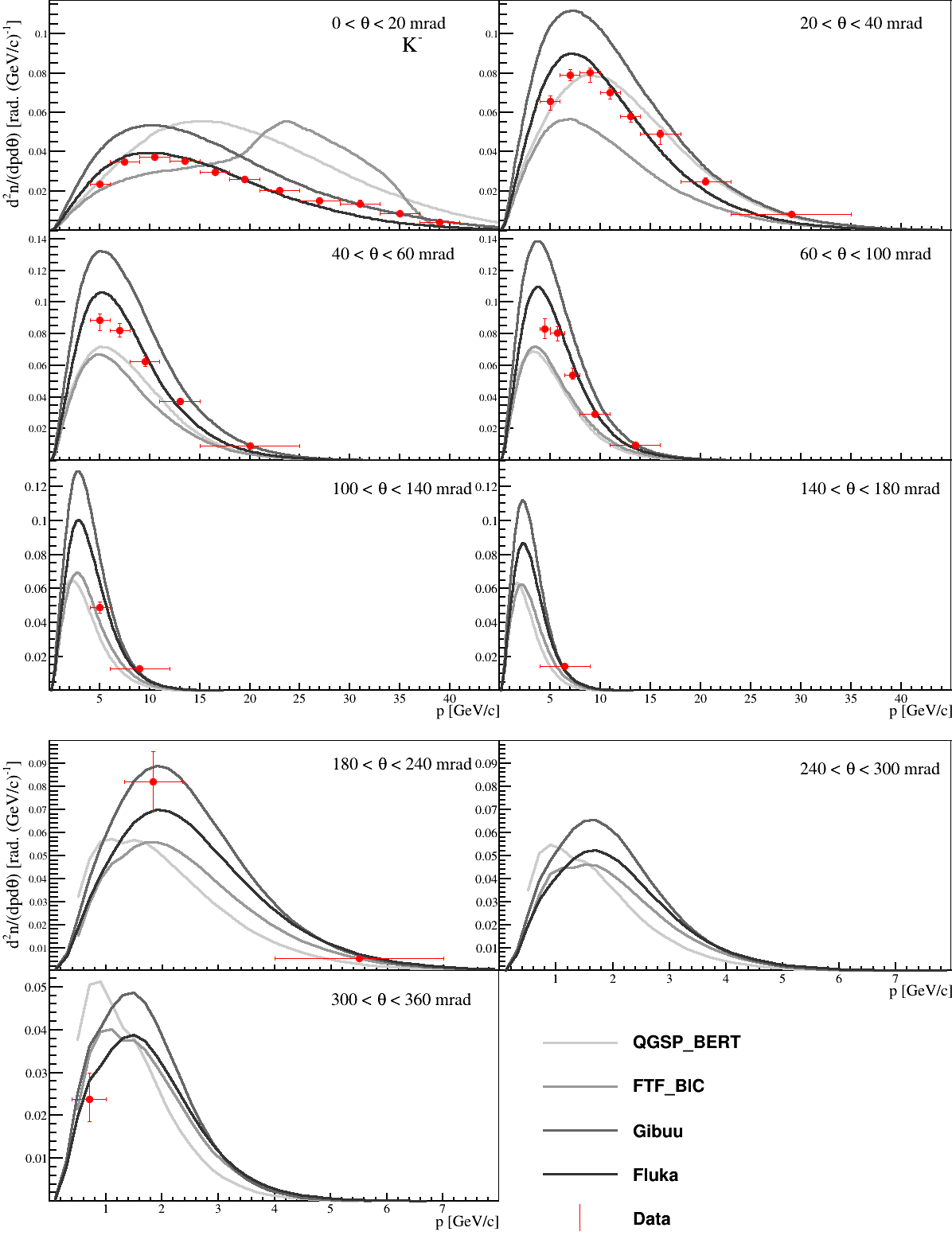}
\caption[\km spectra]{\km multiplicity spectra from \piC interactions are shown. The error bars represent total uncertainties except for the normalization uncertainty. The results are compared to the predictions of the GEANT4 physics lists: QGSP\_BERT and FTF\_BIC as well as GiBUU2019 and FLUKA2011.
}
\label{fig:KM_PiC60_1D}
\end{center}
\end{figure*}

\begin{figure*}[p]
\begin{center}
\includegraphics[width=0.85\textwidth]{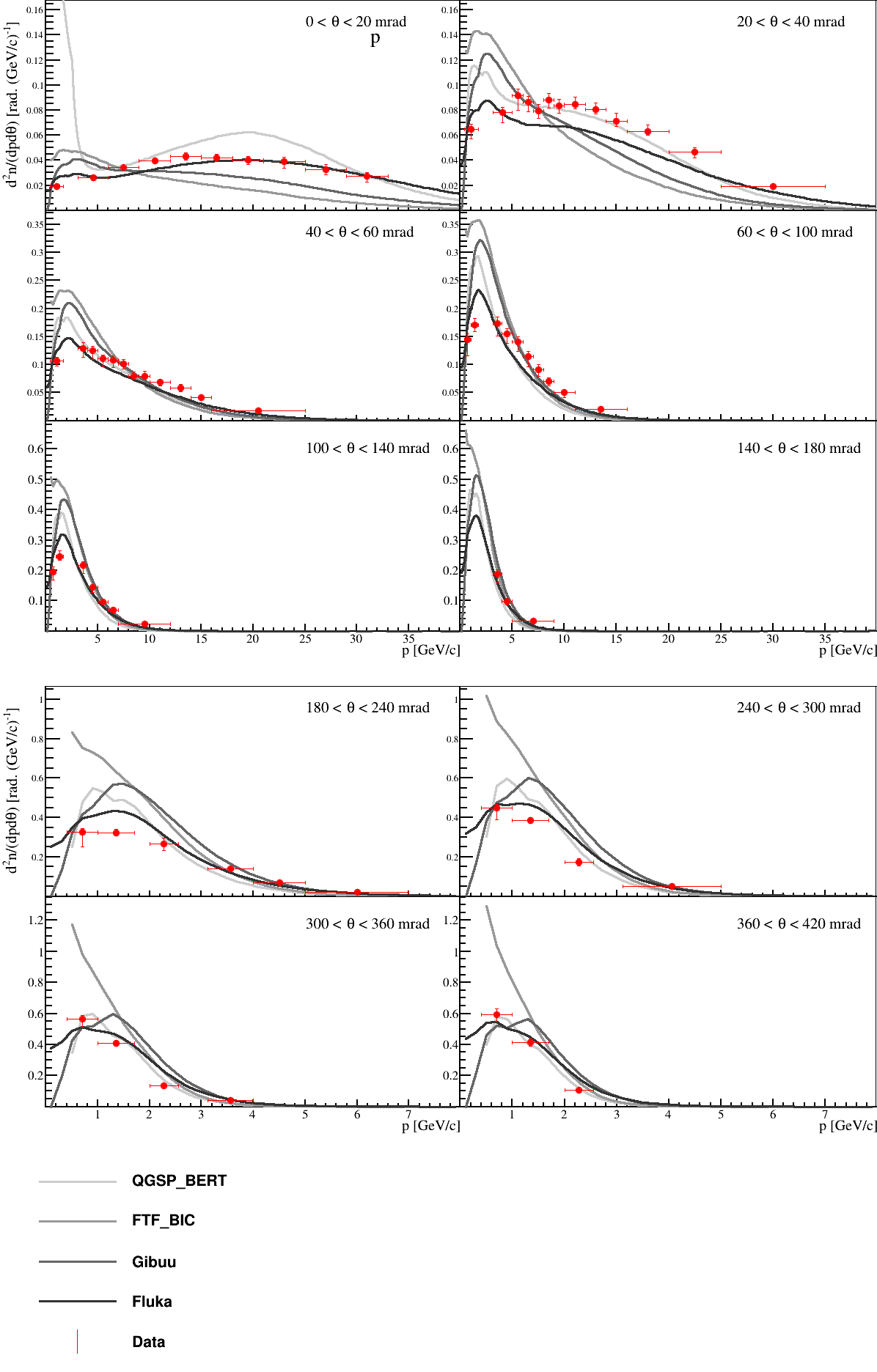}
\caption[Proton spectra]{Proton multiplicity spectra from \piC interactions are shown. The error bars represent total uncertainties except for the normalization uncertainty. The results are compared to the predictions of the GEANT4 physics lists: QGSP\_BERT and FTF\_BIC as well as GiBUU2019 and FLUKA2011.
}
\label{fig:PP_PiC60_1D}
\end{center}
\end{figure*}

Multiplicity spectra obtained for charged pions, charged kaons and protons in \piC interactions are shown in Figures~\ref{fig:PiP_PiC60_1D} through \ref{fig:PP_PiC60_1D}. The results are compared to the predictions of the GEANT4 physics lists: QGSP\_BERT and FTF\_BIC as well as GiBUU2019 and FLUKA2011. In general, no single model provides a good description of the charged hadron spectra for all particle species and for the full phase space, but FLUKA2011 seems to provide the best overall description. However, it should be noted that in the first few angular bins, an important region for neutrino beams, FLUKA2011 slightly over-predicts the production of \pip. Tables in Ref. \cite{edmsTables} present the numerical values of the multiplicity measurements of charged pions, charged kaons and protons along with statistical, systematic and total uncertainties for each kinematic bin analyzed. The normalization uncertainty of $\pm^{2}_{1}$\% is not included in the values of the uncertainties shown in these tables but should be attributed to the multiplicity spectra of all hadron species analyzed.


\begin{figure*}[p]
\begin{center}
\includegraphics[width=0.89\textwidth]{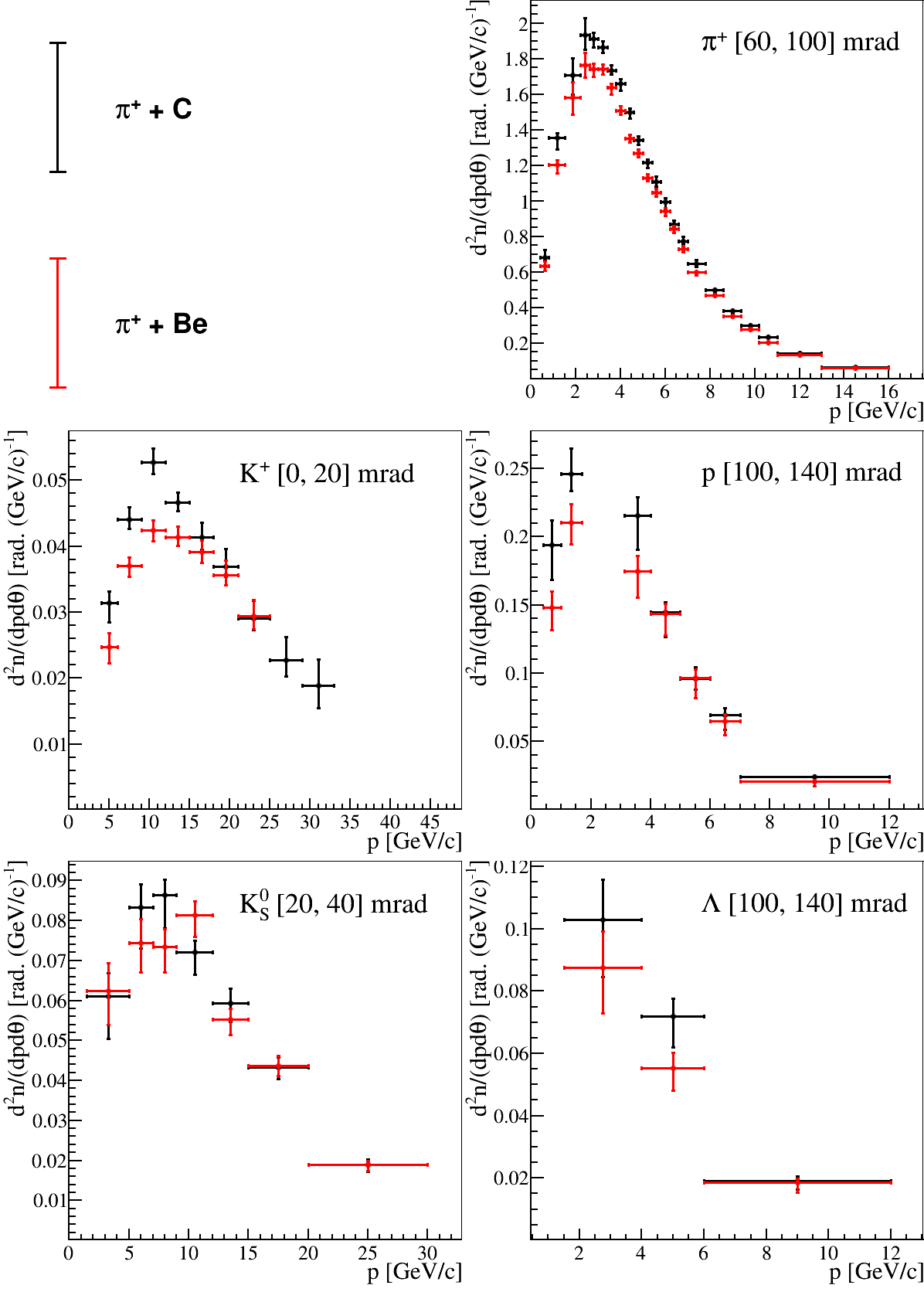}
\caption[Selected \piC \piBe comparison]{Measurements of spectra from \piC and \piBe interactions are shown for produced \pip, \kp, proton, \kos and \lam for select representative angular bins. The error bars represent total uncertainties except for the normalization uncertainty.
}
\label{fig:selectedDataComp}
\end{center}
\end{figure*}

Measurements of spectra of produced \pip, \kp, proton, \kos and \lam from interactions of \piBe are shown in comparison to the results for interactions of \piC for representative angular bins in Figure~\ref{fig:selectedDataComp}. The spectra are largely similar. The most notable difference in the spectra is that the multiplicities tend to be lower in the regions of low momentum and high production angle in interactions of \piBe. The full set of comparisons between the spectra results of \piBe and \piC is presented in Ref. \cite{edmsTables}.

\section{Summary and Conclusions}
\label{sec:summary}

In summary, hadron production was studied in interactions of \piC and \piBe. For both of these reactions, the integrated production and inelastic cross sections were measured. Furthermore, differential cross sections were measured for produced \pip, \pim, \kp, \km, protons, \kos, \lam and \alam. The inelastic cross sections measurements are the first to be made at a beam momentum of 60 \GeVc. The production cross section of interactions of \piBe was measured for the first time as well. The differential cross sections were measured for the first time at this beam momentum scale, and compared to previous measurements at lower beam momenta, a larger kinematic phase space and more particle species were studied. These results will enable neutrino flux predictions to be constrained in neutrino experiments using the NuMI beam and future neutrino beam at LBNF. Specifically, these results can be used to reduce the uncertainties associated with secondary interactions of pions in the carbon targets and the beryllium elements in these beam lines.  

\section*{Acknowledgments}
 We would like to thank the CERN EP, BE, HSE and EN Departments for the strong support of NA61/SHINE. We are grateful to the FLUKA team for their help in producing our model comparisons.

This work was supported by
the Hungarian Scientific Research Fund (grant NKFIH 123842\slash123959),
the Polish Ministry of Science
and Higher Education (grants 667\slash N-CERN\slash2010\slash0,
NN\,202\,48\,4339 and NN\,202\,23\,1837), the National Science Centre Poland (grants~2011\slash03\slash N\slash ST2\slash03691,
2013\slash11\slash N\slash ST2\slash03879, 2014\slash13\slash N\slash
ST2\slash02565, 2014\slash14\slash E\slash ST2\slash00018,
2014\slash15\slash B\slash ST2\slash02537 and
2015\slash18\slash M\slash ST2\slash00125, 2015\slash 19\slash N\slash ST2 \slash01689, 2016\slash23\slash B\slash ST2\slash00692, 2017\slash 25\slash N\slash ST2\slash 02575, 2018\slash 30 \slash A \slash ST2 \slash 00226),
the Russian Science Foundation, grant 16-12-10176,
the Russian Academy of Science and the
Russian Foundation for Basic Research (grants 08-02-00018, 09-02-00664
and 12-02-91503-CERN), the Ministry of Science and
Education of the Russian Federation, grant No.\ 3.3380.2017\slash4.6,
 the National Research Nuclear
University MEPhI in the framework of the Russian Academic Excellence
Project (contract No.\ 02.a03.21.0005, 27.08.2013),
the Ministry of Education, Culture, Sports,
Science and Tech\-no\-lo\-gy, Japan, Grant-in-Aid for Sci\-en\-ti\-fic
Research (grants 18071005, 19034011, 19740162, 20740160 and 20039012),
the German Research Foundation (grant GA\,1480/2-2), the
Bulgarian Nuclear Regulatory Agency and the Joint Institute for
Nuclear Research, Dubna (bilateral contract No. 4799-1-18\slash 20),
Bulgarian National Science Fund (grant DN08/11), Ministry of Education
and Science of the Republic of Serbia (grant OI171002), Swiss
Nationalfonds Foundation (grant 200020\-117913/1), ETH Research Grant
TH-01\,07-3 and the U.S.\ Department of Energy.

\bibliographystyle{na61Utphys}
\bibliography{references} % Produces the bibliography via BibTeX.

\providecommand{\href}[2]{#2}\begingroup\raggedright\begin{thebibliography}{10}

\bibitem{na61detector}
N.~Abgrall {\em et~al.}, {[NA61} Collab.]
  \href{http://dx.doi.org/10.1088/1748-0221/9/06/P06005}{{\em JINST} {\bfseries
  9} (2014) P06005},
\href{http://arxiv.org/abs/1401.4699}{{\ttfamily arXiv:1401.4699
  [physics.ins-det]}}.
%%CITATION = ARXIV:1401.4699;%%.

\bibitem{NA612007Pi}
N.~Abgrall {\em et~al.}, {[NA61/SHINE} Collab.]
  \href{http://dx.doi.org/10.1103/PhysRevC.84.034604}{{\em Phys. Rev.}
  {\bfseries C84} (2011) 034604},
\href{http://arxiv.org/abs/1102.0983}{{\ttfamily arXiv:1102.0983 [hep-ex]}}.
%%CITATION = ARXIV:1102.0983;%%.

\bibitem{NA612007K}
N.~Abgrall {\em et~al.}, {[NA61/SHINE} Collab.]
  \href{http://dx.doi.org/10.1103/PhysRevC.85.035210}{{\em Phys. Rev. C}
  {\bfseries 85} (Mar, 2012) 035210}.

\bibitem{NA612007Neutrals}
N.~Abgrall {\em et~al.}, {[NA61/SHINE} Collab.]
  \href{http://dx.doi.org/10.1103/PhysRevC.89.025205}{{\em Phys. Rev. C}
  {\bfseries 89} (Feb, 2014) 025205}.

\bibitem{na61_t2k_thin}
N.~Abgrall {\em et~al.}, {[NA61/SHINE} Collab.]
  \href{http://dx.doi.org/10.1140/epjc/s10052-016-3898-y}{{\em Eur. Phys. J.}
  {\bfseries C76} no.~2, (2016) 84},
\href{http://arxiv.org/abs/1510.02703}{{\ttfamily arXiv:1510.02703 [hep-ex]}}.
%%CITATION = ARXIV:1510.02703;%%.

\bibitem{numi_tdr}
K.~Anderson {\em et~al.}, ``{The NuMI Facility Technical Design Report}.''
  Fermilab-design-1998-01, fermilab-tm-2406, 1998.

\bibitem{dune_physics}
R.~Acciarri {\em et~al.}, {[DUNE} Collab.]
\href{http://arxiv.org/abs/1512.06148}{{\ttfamily arXiv:1512.06148
  [physics.ins-det]}}.
%%CITATION = ARXIV:1512.06148;%%.

\bibitem{dune_idr}
B.~Abi {\em et~al.}, {[DUNE} Collab.]
\href{http://arxiv.org/abs/1807.10334}{{\ttfamily arXiv:1807.10334
  [physics.ins-det]}}.
%%CITATION = ARXIV:1807.10334;%%.

\bibitem{ichep2018}
H.~Schellman, {[DUNE/LBNF} Collab.], ``{The LBNF Neutrino Beam},'' in {\em talk
  at the 39th International Conference on High Energy Physics (ICHEP2018),
  Seoul, Korea}.
\newblock 2018.

\bibitem{nuint2017}
A.~Bashyal, {[DUNE} Collab.], ``{Neutrino Flux Prediction for DUNE},'' in {\em
  talk at the 11th International Workshop on Neutrino-Nucleus Scattering in the
  Few-GeV Region (NuINT2017), Toronto, Canada}.
\newblock 2017.

\bibitem{fields_beyond2020}
L.~Fields, ``{LBNF Hadron Production Needs and Plans},'' in {\em talk at the
  NA61/SHINE Beyond 2020 Workshop, Geneva, Switzerland}.
\newblock 2017.

\bibitem{na49_pC}
C.~Alt {\em et~al.}, {[NA49} Collab.]
  \href{http://dx.doi.org/10.1140/epjc/s10052-006-0165-7}{{\em Eur. Phys. J.}
  {\bfseries C49} (2007) 897--917},
\href{http://arxiv.org/abs/hep-ex/0606028}{{\ttfamily arXiv:hep-ex/0606028
  [hep-ex]}}.
%%CITATION = HEP-EX/0606028;%%.

\bibitem{harppic12}
M.~G. Catanesi {\em et~al.}, {[HARP} Collab.]
  \href{http://dx.doi.org/10.1016/j.astropartphys.2008.02.002}{{\em Astropart.
  Phys.} {\bfseries 29} (2008) 257--281},
\href{http://arxiv.org/abs/0802.0657}{{\ttfamily arXiv:0802.0657 [astro-ph]}}.
%%CITATION = ARXIV:0802.0657;%%.

\bibitem{cedar}
C.~Bovet, S.~Milner, and A.~Placci
\href{http://dx.doi.org/10.1109/TNS.1978.4329375}{{\em IEEE Trans. Nucl. Sci.}
  {\bfseries 25} (1978) 572--576}.
%%CITATION = IETNA,25,572;%%.

\bibitem{cedar82}
C.~Bovet, R.~Maleyran, L.~Piemontese, A.~Placci, and M.~Placidi
{\em CERN-82-13, CERN-YELLOW-82-13} (1982) .
%%CITATION = CERN-82-13;%%.

\bibitem{na61_prod_cross}
A.~Aduszkiewicz {\em et~al.}, {[NA61/SHINE} Collab.]
  \href{http://dx.doi.org/10.1103/PhysRevD.98.052001}{{\em Phys. Rev.}
  {\bfseries D98} no.~5, (2018) 052001},
\href{http://arxiv.org/abs/1805.04546}{{\ttfamily arXiv:1805.04546 [hep-ex]}}.
%%CITATION = ARXIV:1805.04546;%%.

\bibitem{Agostinelli:2002hh}
S.~Agostinelli {\em et~al.}, {[GEANT4} Collab.]
\href{http://dx.doi.org/10.1016/S0168-9002(03)01368-8}{{\em Nucl. Instrum.
  Meth.} {\bfseries A506} (2003) 250--303}.
%%CITATION = NUIMA,A506,250;%%.

\bibitem{Allison:2006ve}
J.~Allison {\em et~al.}
\href{http://dx.doi.org/10.1109/TNS.2006.869826}{{\em IEEE Trans. Nucl. Sci.}
  {\bfseries 53} (2006) 270}.
%%CITATION = IETNA,53,270;%%.

\bibitem{Allison:2016lfl}
J.~Allison {\em et~al.}
\href{http://dx.doi.org/10.1016/j.nima.2016.06.125}{{\em Nucl. Instrum. Meth.}
  {\bfseries A835} (2016) 186--225}.
%%CITATION = NUIMA,A835,186;%%.

\bibitem{nikos}
N.~Charitonidis, ``{Muon Population in NA61}.''
  \url{https://edms.cern.ch/ui/file/1909492/1/Muons_NA61.pdf}, Feb, 2018.
\newblock CERN-EDMS-1909492.

\bibitem{Nikos2}
N.~Charitonidis, ``{Positron Population in NA61},'' 2019.
\newblock Private communication.

\bibitem{Carroll}
A.~Carroll {\em et~al.}
\href{http://dx.doi.org/10.1016/0370-2693(79)90226-0}{{\em Phys.\ Lett.}
  {\bfseries B80} (1979) 319}.
%%CITATION = PHLTA,B80,319;%%.

\bibitem{Denisov:1973zv}
S.~P. Denisov, S.~V. Donskov, {\relax Yu}.~P. Gorin, R.~N. Krasnokutsky, A.~I.
  Petrukhin, {\relax Yu}.~D. Prokoshkin, and D.~A. Stoyanova
\href{http://dx.doi.org/10.1016/0550-3213(73)90351-9}{{\em Nucl. Phys.}
  {\bfseries B61} (1973) 62--76}.
%%CITATION = NUPHA,B61,62;%%.

\bibitem{PhysRevD.98.030001}
M.~Tanabashi {\em et~al.}, {[Particle Data Group} Collab.]
  \href{http://dx.doi.org/10.1103/PhysRevD.98.030001}{{\em Phys. Rev. D}
  {\bfseries 98} (Aug, 2018) 030001}.

\bibitem{edmsTables}
S.~R. Johnson {\em et~al.}, ``{Tables with numerical results for paper on
  hadron production from 2016 pion data}.''
  \url{https://edms.cern.ch/document/2215444}, 2019.
\newblock CERN-EDMS-2215444.

\bibitem{Buss:2011mx}
O.~Buss, T.~Gaitanos, K.~Gallmeister, H.~van Hees, M.~Kaskulov, O.~Lalakulich,
  A.~B. Larionov, T.~Leitner, J.~Weil, and U.~Mosel
  \href{http://dx.doi.org/10.1016/j.physrep.2011.12.001}{{\em Phys. Rept.}
  {\bfseries 512} (2012) 1--124},
\href{http://arxiv.org/abs/1106.1344}{{\ttfamily arXiv:1106.1344 [hep-ph]}}.
%%CITATION = ARXIV:1106.1344;%%.

\bibitem{Battistoni:2015epi}
G.~Battistoni {\em et~al.}, ``{Overview of the FLUKA code},'' 2015.

\bibitem{Bohlen:2014buj}
T.~T. Bohlen, F.~Cerutti, M.~P.~W. Chin, A.~Fasso, A.~Ferrari, P.~G. Ortega,
  A.~Mairani, P.~R. Sala, G.~Smirnov, and V.~Vlachoudis
\href{http://dx.doi.org/10.1016/j.nds.2014.07.049}{{\em Nucl. Data Sheets}
  {\bfseries 120} (2014) 211--214}.
%%CITATION = NDTSB,120,211;%%.

\bibitem{Ferrari:2005zk}
A.~Ferrari, P.~R. Sala, A.~Fasso, and J.~Ranft, ``{FLUKA: A multi-particle
  transport code},'' 2005.

\end{thebibliography}\endgroup

\end{document}